\documentclass[submission, Phys]{new}

\usepackage[T1]{fontenc}
\usepackage[latin9]{luainputenc}
\usepackage[T1]{fontenc}
\usepackage[latin9]{luainputenc}
\usepackage{amsmath,bm}
\usepackage{titlesec}
\usepackage{blindtext}
\usepackage{amssymb}
\usepackage{graphicx}
\usepackage{ulem}   
\usepackage{xcolor}
\usepackage{bbold}
\usepackage{comment}
\hypersetup{colorlinks,allcolors=black}
\usepackage{amssymb}
\usepackage{gensymb}
\usepackage{float}
\usepackage[english]{babel}
\usepackage{tabularx,graphicx}
\usepackage{epstopdf}
\usepackage{latexsym}
\usepackage{color, colortbl}
\usepackage{psfrag}
\usepackage{bbm}
\usepackage{bm}
\usepackage{titlesec}
\usepackage{dsfont}
\usepackage{feynmp}
\usepackage{slashed}
\usepackage{multirow}
\usepackage{tocloft}

\renewcommand{\vec}[1]{\boldsymbol{#1}}
\def \nn {\nonumber}

\def \uar{{\uparrow}}

\def \k {{\vec k}}

\def \ve {\varepsilon}
\def \r {{\vec r}}

\def \q {{\vec q}}

\def \l {\ell}
\def \ve {\varepsilon}

\def \l {\ell}

\def \beq {\begin{eqnarray}}
\def \eeq {\end{eqnarray}}
\def \tn {\textnormal}

\def \k {{\vec k}}

\def \la{\langle}
\def \ra{\rangle}

\def \hk {H_{\tn{kin}}}
\def \hi {H_{\tn{int}}}
\def \kx {K_{\tn{xx}}}
\def \kxx {K_{\tn{xx}}^{\tn{eff}}}
\def \kxxnaive {K_{\tn{xx}}^{\tn{naive}}}
\def \J {J^{\tn{eff}}}
\def \cx {\chi_{\tn{xx}}}
\def \cxx {\chi_{\tn{xx}}^{\tn{eff}}}

\def \heff {\mathcal{H}_{\tn{eff}}}
\def \gs {|\tn{GS}\rangle}
\def \g0 {|\tn{GS}_0\rangle}
\def \ef {E_0^{\tn{eff}}}

\begin{document}
\begin{center}{\Large \textbf{Upper bounds on superconducting and excitonic phase-stiffness for interacting isolated narrow bands}}\end{center} 
\begin{center}
Dan Mao,
Debanjan Chowdhury\textsuperscript{*}
\end{center}
\begin{center}
Department of Physics, Cornell University, Ithaca, New York 14853, USA.
\\
* debanjanchowdhury@cornell.edu\\
\end{center}
\begin{center}
\today
\end{center}

\section*{\begin{center}Abstract\end{center}}
{\bf
Inspired by the discovery of superconductivity in moir\'e materials with isolated narrow bandwidth electronic bands, here we analyze critically the question of what is the maximum attainable $T_c$ in interacting flat-band systems. We focus specifically on the low-energy effective theory, where the density-density interactions are projected to the set of partially-filled flat bands. The resulting problem is inherently non-perturbative, where the standard mean-field approximation is not applicable. Here we develop further our recent Schrieffer-Wolff transformation based approach (\href{https://www.pnas.org/doi/10.1073/pnas.2217816120}{\color{blue} PNAS, 120 (11), e2217816120 (2023)}) to compute the effective electromagnetic response and the superconducting phase-stiffness in terms of ``projected'' gauge-transformations and extend the formalism to compute the stiffness for excitonic superfluids. Importantly, our method requires neither any ``wannierization'' for the narrow bands of interest, regardless of their (non-)topological character, nor any knowledge of an underlying pairing symmetry, and can be set up directly in momentum-space. We use this formalism to derive upper bounds on the phase-stiffness for sign-problem-free models, where their values are known independently from numerically exact quantum Monte-Carlo computations. We also illustrate the analytical structure of these bounds for the superconducting and excitonic phase-stiffness for perfectly flat-bands with Landau-level-like wavefunctions.   
}

\clearpage
\hrule
\tableofcontents
\vspace{0.5cm}
\hrule
\section{Introduction}

Making precise theoretical statements about the superconducting transition temperature ($T_c$) --- an inherently non-universal dimensionful quantity --- in the regime of strong interactions, where Bardeen-Cooper-Schrieffer (BCS) theory is {\it a priori} inapplicable is an insurmountable task. The problem is made especially difficult by the possibility of the interactions driving a panoply of competing orders, which can drive $T_c$ to be vanishingly small. On the other hand, addressing the fundamental question of what is the highest achievable $T_c$ for a given microscopic electronic model is of self-evident interest. Given any electronic system with a few characteristic (``bare'') energy scales, such as the bandwidth, $W$, typical interaction strength, $U$, Debye frequency, $\omega_D$, and so on, it is reasonable to expect that $T_c\leq\tn{max}(W,U,\omega_D,...)$. In reality, the actual $T_c$ is typically much smaller than any of these energy scales. The key difficulty in trying to engineer higher temperature superconductors is that optimizing the pairing (gap) scale often comes at the expense of the phase-coherence scale \cite{EK95,BOK,MR10}. 

The discovery of superconductivity in moir\'e materials \cite{Cao2018,AY19,Efetov19,SNP20,PK21,park2021tunable} has ushered a new era, where the enhancement in the pairing gap-scale and the phase-coherence scale do not necessarily compete against one another. The common ingredient across these materials is a set of partially filled isolated bands with a narrow bandwidth and an interaction strength that is comparable to, or larger than, the bandwidth. 
It is conceivable that there exist non-perturbative regimes (e.g. in the flat-band limit) beyond the conventional BCS picture of superconductivity, where the interaction-scale enhances both the pairing-gap scale and the phase-coherence scale. However, addressing this regime should not rely on an unjustified application of ``weak-coupling'' BCS theory and its various stronger coupling extensions (that rely on an adiabatic connection to the weak-coupling regime).  In the absence of controlled analytical methods in this new non-perturbative regime, a number of numerically exact quantum Monte-Carlo (QMC) studies have demonstrated unambiguously that interacting, nearly flat bands can support superconductivity \cite{hofmann2020superconductivity,peri2021fragile,Bernevig21,Zhang2021,hofmann2022superconductivity}; see also  Refs.~\cite{TBGV,herzog2022many} for some exactly soluble models. However, infinitesimal perturbations can destroy superconductivity and induce competing orders \cite{hofmann2020superconductivity,hofmann2022superconductivity}; predicting this competition reliably lies well beyond the scope of any weak-coupling approach. 

The goal of this paper is to address the following fundamental question regarding superconductivity: {\it Given a microscopic electronic model with a single-particle spectrum involving multiple isolated bands and density-density interactions, what is the highest $T_c$ that can be achieved if the low-energy physics is determined by only a subset of partially-filled bands?} We are interested in problems in two-dimensional systems, where $T_c$ is limited by phase fluctuations and the transition is determined by the criterion $T_c=\pi D_s(T_c^{-})/2$  \cite{nelson1977universal}, where $D_s$ is the superconducting phase-stiffness. Therefore, the above question is equivalent to addressing what bounds $D_s$ from above at low energies. While the question is sufficiently general and relevant for numerous electronic solids, we will primarily focus on examples involving isolated bands with a narrow bandwidth, as is relevant e.g. for  moir\'e materials. Relatedly, in problems involving isolated, nearly flat bands where there is a tendency towards the formation of excitonic superfluidity, we will address the question of what is the largest possible excitonic phase-stiffness, without invoking any Hartree-Fock approximations.

The superconducting phase-stiffness is related to a diamagnetic, i.e. an electromagnetic response. In order to compute such response functions for the low-energy theory, we have to couple an external (probe) vector potential to the low-energy ``active'' degrees of freedom. The usual way to do this for electronic (tight-binding) lattice models, where the information for all of the UV degrees of freedom are retained is to carry out a Peierls'-type substitution. However, for models projected to topological bands with a finite Chern number (where a tight-binding description does not exist), or continuum-type models defined directly in momentum-space which are often the starting point for describing moir\'e systems, the correct procedure is {\it a priori} unclear. Moreover, computing the effective low-energy electromagnetic response requires one to systematically integrate out the effects of high-energy (remote) bands to obtain a theory associated solely with the active bands of interest. We focus on some of these subtleties and clarify various misconceptions regarding what the effective low-energy electromagnetic response for the models of interest should entail using the bandstructure of twisted-bilayer graphene (TBG) as an illustrative example in the remainder of this section. We will address the question specifically for TBG and related moir\'e materials in a future publication \cite{DMJFMVDC}.

\begin{figure}
    \centering
    \includegraphics[width=0.98\textwidth]{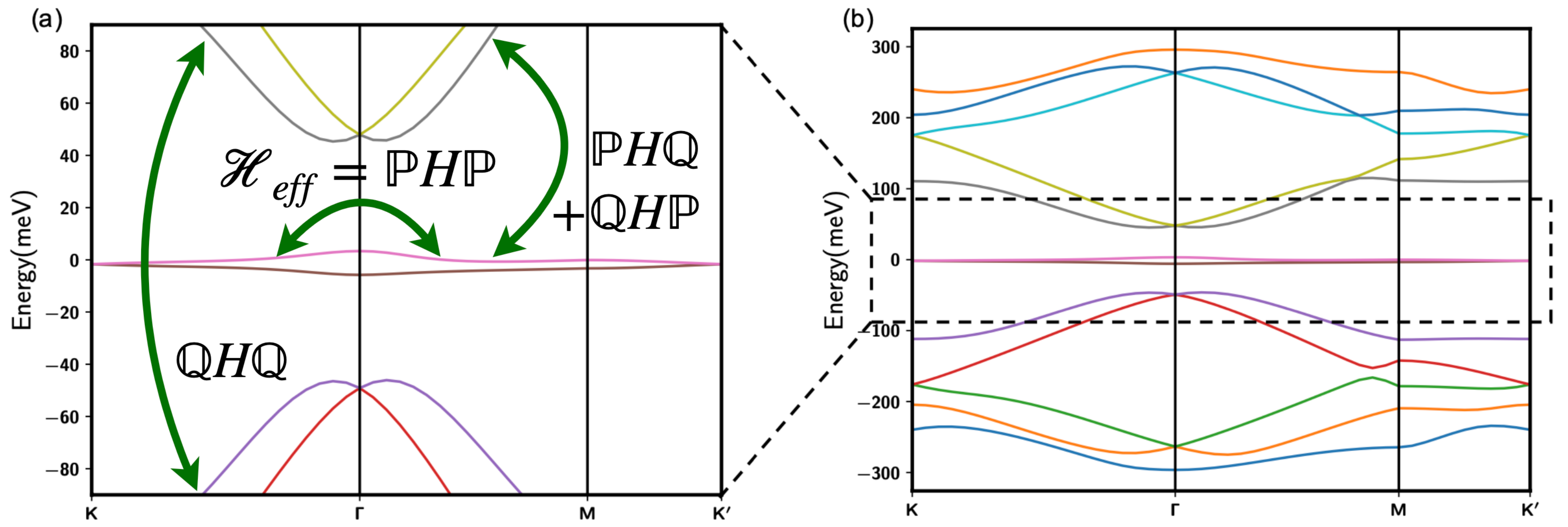}
    \caption{A typical non-interacting dispersion obtained for twisted bilayer graphene (TBG) near magic-angle ($\theta=1.12^\circ$) using a continuum-model \cite{BM11} with ratio of interlayer-hopping parameters, $\kappa = w_{\tn{AA}}/w_{\tn{AB}} = 0.7$. A specific cut along a high-symmetry direction is shown in the moir\'e Brillouin-zone. A goal of this paper is to obtain an estimate of $D_s$ for simpler models involving (a) isolated nearly ``flat'' bands at low energy that are well isolated from the dispersive remote bands, as shown in the zoomed-out bandstructure in panel (b), taking into account the contribution of the projected density-density interactions, $\heff = \mathbb{P} H\mathbb{P}$. The contributions involving the other projections $\mathbb{Q} H\mathbb{Q}$ and $(\mathbb{Q}H\mathbb{P}+\mathbb{P}H\mathbb{Q})$ are also shown; see Eqn.~\ref{projection}. As a matter of principle, the electromagnetic response associated with {\it any} subset of bands (e.g. either those in (a) or (b)) derived from the original band for untwisted, decoupled layers of graphene {\it must} necessarily include a contribution from the projected interactions that couple to the vector potential.}
    \label{fig:many_TBG_bands}
\end{figure}

For the low-energy electromagnetic response for problems of interest to us, working with the entire single-particle spectrum that contains {\it all} electronic bands is not desirable (and often unnecessary). To illustrate this point, it is best to take the case of TBG as a concrete example. Firstly, it is possible to start with a microscopic tight-binding model for the two sheets of graphene and include the inter-layer hoppings to arrive at a microscopic tight-binding model for TBG. Clearly, this includes all the information about the individual atoms inside the large moir\'e unit cell in TBG. However, this UV Hamiltonian has an electronic bandwidth set by the bandwidth of original graphene (i.e. $O($eV$)$), albeit with many folded bands, including the isolated flat-band of interest to us. If one starts with the  (screened) Coulomb interaction, it is possible to obtain the electromagnetic response associated with the full Hamiltonian, which necessarily includes the contribution from {\it all} the bands. In this approach, the density-density interaction does not couple to the external vector potential, but the microscopic current operator includes inter and intra-band contributions from the entire spectrum.  On the other hand, if we are interested in the low-energy response associated with just the isolated ``flat'' bands (Fig.~\ref{fig:many_TBG_bands}a), or even a subset of the ``remote'' bands (Fig.~\ref{fig:many_TBG_bands}b), it is no longer legitimate to ignore the contribution of the interaction terms to the electromagnetic response. As will be demonstrated in this paper, the act of {\it projecting} the interactions to the ``target'' set of bands (which can include a subset of the remote bands) and {\it integrating-out} the higher-energy bands leads to non-trivial interaction-induced contributions to the electromagnetic response. Note that this statement does not rely on the non-trivial (fragile) topological character associated with the bands in twisted bilayer graphene \cite{LZ1,LZ2,BAB1,BJY}, and these considerations are relevant even for topologically trivial bands. This manuscript will extend our recent approach \cite{DMDC1} and outline a clear procedure to: (i) couple a probe vector-potential to the active charge-carrying degrees of freedom in the low-energy effective theory, without requiring any ``wannierized'' description, to obtain the diamagnetic response, and (ii) integrate out the contributions from remote bands via a Schrieffer-Wolff (or a similar approach perturbative in the inverse gap to remote-bands) transformation. One of the main objectives of the present manuscript is to apply the resulting formalism to interacting flat-band models to derive upper bounds on $D_s$, and compare against numerically exact results for the same obtained from QMC. With a simple modification, we will also be able to use the same basic setup to compute the ``exciton phase-stiffness'' for two-component flat-band systems in the strongly interacting regime. To gain additional insight into these interacting correlation functions, we will evaluate them explicitly for one of the best studied examples of a topological flat-band, namely the lowest Landau-level.

The remainder of this article is organized as follows: In Sec.~\ref{sec:prelim} we formulate the basic setup of the problem and discuss the simplified limit of a multi-band model of non-interacting electrons as an illustrative example. In Sec.~\ref{sec:theory} we introduce two complementary methods to evaluate the transverse electromagnetic response for density-density interactions projected to isolated flat-bands. Sec.~\ref{sec:bounds} applies the framework to two concrete microscopic models  involving topological bands as well as bands with fragile topology, where sign-problem free QMC computations have obtained the actual $D_s$. In Sec.~\ref{sec:exciton} we extend the above framework to compute the stiffness associated with an excitonic superfluid, where the probe gauge-field is distinct from the physical gauge-field. Sec.~\ref{sec:ll} focuses specifically on applications of the above methods to lowest Landau-level-like wavefunctions with a (non-)uniform distribution of Berry curvature. We conclude with a discussion of the relationship of this work to earlier works based on application of BCS mean-field theory, focusing specifically on the role of the Fubini-Study metric, and outline some of the open questions in Sec.~\ref{sec:BCS}.

\section{Preliminaries}
\label{sec:prelim}
In this section, we will summarize the scope of our discussion in terms of the family of models that we analyze in this paper and the computational setup that forms the basis of much of our analysis. We also discuss the example of non-interacting electrons with an energy spectrum involving isolated bands as a warm-up exercise to highlight the need for general caution while using minimal-coupling and projection to compute the electromagnetic response. 

\subsection{Electromagnetic response functions for low-energy effective theory}
\label{sec:EM}

We begin this section by recalling the paramagnetic and diamagnetic response functions associated with electronic models, and the three distinct limits associated with how the (transverse) momentum and frequency are taken to zero. As is well known, these limits correspond to very different physical settings \cite{scalapino1993insulator}. Starting with the full  paramagnetic current susceptibility, $\chi_{\mu\nu}(\q, \omega)$, and the diamagnetic response, $K_{\mu\nu}$, the distinct response functions are given by,
\begin{subequations}
\beq
\tn{Drude weight:}~~~~D&=& \frac{e^2}{4}\bigg[\la K_{\tn{xx}}\ra - \chi_{\tn{xx}}(\q = 0, \omega\rightarrow 0)\bigg], \\
\tn{Superfluid stiffness:}~~~~ D_s &=& \frac{e^2}{4}\bigg[\la  K_{\tn{xx}} \ra - \chi_{\tn{xx}}(q_x = 0, q_y \rightarrow 0 ,\omega = 0)\bigg], \label{Aresponse} \\
\tn{Longitudinal response:}~~~~ \la  K_{\tn{xx}} \ra &-& \chi_{\tn{xx}}(q_x \rightarrow 0, q_y = 0 ,\omega = 0)  = 0. 
\eeq
\end{subequations}
Here, gauge-invariance requires that the total longitudinal response vanish.

In general, we are interested in computing the electromagnetic response (specifically, $D_s$) for any electronic lattice Hamiltonian involving a set of multiple orbitals (including spin), where the electrons interact only via density-density interactions as,
\begin{subequations}
\beq
H &=& \hk + \hi, \label{ham1}\\
\hk &=& \sum_{\substack{\r,\r'\\ \alpha,\alpha'}} t_{\alpha\alpha'}(\r-\r') c^\dagger_{\r\alpha} c^{\phantom\dagger}_{\r'\alpha'} - \mu N,\label{ham2}\\
\hi &=& \sum_{\r,\r'} V(\r-\r')~ n_\r n_{\r'}.\label{ham3}
\eeq
\end{subequations}
Here $c_{\r\alpha},~c^\dagger_{\r\alpha}$ denote microscopic electronic operators at site $\r$ with orbital/spin labels $\alpha,~\alpha'$, with a matrix $t_{\alpha\alpha'}(\r-\r')$ determining the full non-interacting bandstructure. The chemical potential, $\mu$, couples to the (global) conserved $U(1)$ density, $N = \sum_{\r}n_\r = \sum_{\r,\alpha} c^\dagger_{\r\alpha} c^{\phantom\dagger}_{\r\alpha}$. In particular, we are interested in the limit where the non-interacting bandwidth of the low energy ``active'' bands $W$ and the interaction $V$ are much smaller than the band-gap $\Delta$ separating the active bands from the remote bands.

We emphasize at the outset that for simplicity we have ignored other types of interaction, including most notably the electron-phonon interaction. This is an important point, since the form of the interaction determines how an external vector potential ($\vec{A}$) couples to the Hamiltonian. For the full Hamiltonian introduced above, $\vec{A}$ couples {\it only} to $\hk$ and not to $\hi$. We can obtain the microscopic current operator and the diamagnetic response starting with $H$ as,
\begin{subequations}
\beq
J_\mu(q_\mu \rightarrow 0) &=&  -i \left[\hat{X}_\mu, H\right], \label{jmu}\\
K_{\mu\nu} &=& -\left[\hat{X}_\mu, \left[\hat{X}_\nu, H\right]\right],~\tn{where} \label{kxx}\\
\hat{X}_\mu &\equiv& \sum_i x_i^\mu c_i^\dag c_i^{\phantom\dagger}
\eeq
\end{subequations}
 is the many-body position operator. Note that the current operator and the diamagnetic term do {\it not} depend on the choice of origin of the position operator since changing the origin results in a constant shift of the position operator, which commutes with the Hamiltonian. Starting with these observables, one can compute the response functions as introduced in Eq.~\ref{Aresponse} in a specific many-body state. It is useful to realize that when dealing with the entire spectrum of $\hk$, nominally $J_\mu$ and $K_{\mu\nu}$ depend {\it explicitly} only on the entries of $t_{\alpha\alpha'}(\r-\r')$ and not on the interactions. However, the dependence on the latter enters implicitly when evaluating the expectation values, $\la K_{\tn{xx}}\ra$ and $\chi_{\tn{xx}}$, in the many-body state.    

The first step in our framework is to obtain the effective Hamiltonian, $\heff[{\vec A}]$, that captures the low-energy physics associated with a subset of isolated bands associated with $\hk$. One of the appealing features is that it will allow us to work exclusively in momentum-space, obviating the need to work with any tight-binding models in real-space. This is especially useful if one is interested, e.g. in dealing with the continuum-type models relevant for moir\'e systems.  

We are interested in computing,
\begin{subequations}
 \beq
        \frac{4 D_s}{e^2} &=& \bigg[\langle \kxx \rangle - \cxx(\omega=q_x=0,q_y \rightarrow 0)\bigg]_{A\rightarrow0}, \label{Ds}\\
       \kxx &=& \frac12 \frac{\delta^2 \heff[A]}{\delta A_\tn{x} \delta A_\tn{x}},~~
        \cxx(\q) = \langle \J_\tn{x}(\q)~\J_\tn{x}(-\q)\rangle,\label{susc}
\label{eq:D}
\eeq
\end{subequations}
where $\kxx$ is the {\it effective} diamagnetic contribution and $\cxx$ is the {\it effective} current susceptibility, with $\J_\tn{x} = -\delta \heff[A]/\delta A_{\tn{x}}$. {\it A priori}, obtaining $\heff[A]$ given the UV Hamiltonian $H$ in Eq.~\ref{ham1} is a non-trivial task. We will describe the procedure to extract this information and the related susceptibilities in Sec.~\ref{sec:theory} below. 
Importantly, there is a ``partial f-sum rule'' that relates $\langle\kxx\rangle$ to the integrated (longitudinal) optical spectral weight associated with $\heff$ \cite{MR21,DMDC1}, 
\begin{equation}
    \int_0^\Lambda d\omega ~\tn{Re} [\sigma^{\tn{eff}}_{\tn{xx}}(\omega)] = \frac{\pi e^2}{2 } \langle \kxx \rangle,
    \label{eq:partial_f_sum}
\end{equation}
where $\Lambda$ is chosen to lie inside the gap to the remote bands.

\subsection{Warm-up example: multi-band model of non-interacting electrons}\label{sec:non-int-multi}

This section provides a brief exposition to some of the subtleties that can arise when adopting different procedures for introducing a vector-potential in a theory with multiple separated electronic bands. Interestingly, these differences can already be seen in the limit where $\hi=0$. To be clear, in the absence of interactions, the models do {\it not} have the possibility of supporting superconductivity (i.e. $D_s=0$) and so we will not make any reference to ``bounds'' on $T_c$. Instead, we will consider the electromagnetic response for $\hk$ in various non-commuting limits of $\{q_x,~q_y,~\omega\}\rightarrow 0$ in the limit of a weak probe gauge-field, $A_x\rightarrow0,~A_y=0$, and analyze the key differences. 

The diamagnetic susceptibility contains three classes of terms,
\beq
        \langle \kx \rangle = \sum_{\k,m} f(\ve_{\k m}) \Big[\partial_{k_x}^2\ve_{\k m}
+ \ve_{\k m} \left(\langle u_{\k,m}| \partial_{k_x}^2 u_{\k,m} \rangle + \langle  \partial_{k_x}^2 u_{\k,m}| u_{\k,m} \rangle\right)\nn\\
+ 2 \langle \partial_{k_x} u_{\k,m}| \hat{h}_\k| \partial_{k_x} u_{\k,m}\rangle\Big],
    \label{eq:k_xx}
\eeq
where $f(\ve_{\k m}) = \langle c_{\k m}^\dag c^{\phantom\dagger}_{\k m}\rangle $ is the Fermi-Dirac distribution and $\hat{h}_\k = \sum_m \ve_{\k m} |u_{\k,m}\rangle \langle u_{\k,m} |$ is the Hamiltonian in momentum space. It is important to realize that ultimately, only the first term in Eq.~\ref{eq:k_xx} contributes to the physical response, once the contributions from $\cx$ are included in the full $D_s$. In fact, the above observation already illustrates that $ \langle  \kx \rangle$ does not provide an ``intrinsic'' response associated with the active bands since it depends on the band-gap between the active band and remote bands. Therefore, in order to obtain an intrinsic response function, one must take into account the contribution from the paramagnetic term $\cx(q,\omega)$.

Evaluating $\cx(\q=0, \omega \rightarrow 0)$ leads to the straightforward result,
\begin{subequations}
\beq
\cx(\q=0, \omega \rightarrow 0) 
    &=& 2 \sum_{\k,m,n} f(\ve_{\k m}) \left[1- f(\ve_{\k n})\right] 
    \times\frac{ \langle u_{\k,m} | \partial_{k_x} \hat{h}_\k |u_{\k,n}\rangle \langle u_{\k,n} | \partial_{k_x} \hat{h}_\k |u_{\k,m}\rangle}{\ve_{\k n} - \ve_{\k m}} \nn \\ \\
    &=& 2 \sum_{\k,m} f(\ve_{\k m}) \Bigg(\langle \partial_{k_x} u_{\k,m} |\hat{h}_\k - \ve_{\k m}| \partial_{k_x} u_{\k,m}\rangle\Bigg).
\label{eq:kxx_0}
\eeq
\end{subequations}
When combined with $\langle \kx\rangle$, this leads to the usual Drude weight,
\begin{equation}
    \frac{4 D}{ e^2} = \sum_{\k,m} f(\ve_{\k m}) \partial_{k_x}^2 \ve_{\k m}.
\end{equation}
At $T=0$, the above quantity only depends on the {\it partially} occupied bands.

The static limit, $\q\rightarrow 0, \omega = 0$, relevant for the superfluid stiffness is fundamentally different from the limit just considered above.  For $q =0$, the only non-vanishing contribution to $\cx$ arises from inter-band terms ($m\neq n$). On the other hand, since $J(\q)$ connects states with different momenta for $\q \neq 0$, intra-band terms with $m=n$ can contribute to $\cx$. This leads to,
\begin{subequations}
\beq
 \cx(\q\rightarrow 0, \omega = 0) 
&=&\lim_{\q\rightarrow0} 2 \sum_{\k,m,n}  f(\ve_{\k_+,m}) \left[1- f(\ve_{\k_-,n})\right] \nn\\
&&\times\frac{ \langle u_{\k_+,m} | \partial_{k_x} \hat{h}_\k |u_{\k_-,n}\rangle \langle u_{\k_-,n} | \partial_{k_x} \hat{h}_\k |u_{\k_+,m}\rangle}{\ve_{\k_-,n} - \ve_{\k_+,m}} \nn \\ \\
&=& \cx(q=0,\omega\rightarrow 0) 
- 2\sum_{\k,m} f(\ve_{\k, m}) \partial_{\epsilon_{\k,m}}f(\ve_{\k,m}) (\partial_{k_x} \ve_{\k,m})^2 \nn\\
&=& \cx(q=0,\omega\rightarrow 0) + \sum_{\k,m} f(\ve_{\k,m}) \partial_{k_x}^2 \ve_{\k,m}, \label{eq:kxx_static}
\eeq
\end{subequations}
where $\k_\pm = \k\pm\q/2$ and the second line is obtained by expanding in a small $\q$. Note that the final result in Eq.\ref{eq:kxx_static} does not depend on how $q \rightarrow 0$, and we obtain $D_s = 0$, as expected for non-interacting electron systems.

An alternative approach towards obtaining $D_s$ is to work directly with the Hamiltonian projected to the active bands. We denote the quantities computed using the projected Hamiltonians with an ``$\overline{~~}$''. A simple minded application of Eq.~\ref{eq:D} leads to unphysical results, as we demonstrate below. For the sake of simplicity, let us consider only one active band, such that
\begin{equation}
\begin{split}
    \langle \overline{\kx} \rangle =\sum_\k f(\ve_\k) \langle u_\k | \partial_{k_x}^2 \overline{h}_\k |u_\k\rangle =\sum_\k f(\ve_\k)\left[\partial_{k_x}^2\ve_{\k} - 2 \ve_\k g_{\tn{xx}}(\k)\right],
\end{split}
\label{eq:K_bar}
\end{equation}
where $\overline{h}_\k = \ve_\k |u_\k\rangle \langle u_\k |$ is the projected Hamiltonian and 
\beq
g_{\tn{xx}}(\k) = \langle \partial_{k_x} u_\k |\left(1-| u_\k \rangle \langle u_\k | \right)|\partial_{k_x} u_\k \rangle
\eeq
is the quantum metric. The paramagnetic terms, $\overline{\cx}(\q=0,\omega\rightarrow 0) = 0$ and $\overline{\cx}(\q\rightarrow 0,\omega= 0) = \sum_\k f(\ve_\k) \partial_{k_x}^2 \ve_\k$. The ``response functions'' obtained from $\overline{\kx}$ and $\overline{\cx}$ are unphysical if we blindly apply Eq.\ref{eq:D}. First of all, they are not intrinsic because $\langle \overline{\kx} \rangle$ depends on the actual value of the energy $\ve_\k$ itself (instead of its derivatives w.r.t $\k$, the bandwidth etc.). Even more surprisingly, we get a non-vanishing longitudinal contribution, which seemingly ``violates'' gauge invariance. This violation can be viewed as the consequence of the {\it non-unitarity} of the induced gauge transformation, when restricted to the projected Hilbert space \cite{DMDC1}. For example, for an infinitesimal gauge transformation generated by  $U_{\delta} = e^{i \vec\delta\cdot\r}$, the transformation restricted to the projected Hilbert space associated with only one active band gives $c_{\k} \rightarrow \mathbb{P} U_{\delta}^\dag c_{\k}U_{\delta} \mathbb{P} = c_{\k+\vec{\delta}} \langle u_{\k}|u_{\k+\vec\delta}\rangle$, where $\mathbb{P}$ is the projector. This restricted transformation is not unitary; for instance if $\vec\delta=\delta \hat{x}$, the measure of the path integral changes as $|\langle u_\k |u_{\k+\vec\delta}\rangle|^2 \approx 1- \delta^2 g_{\tn{xx}}(\k)$, which explains the origin of the extra $g_{\tn{xx}}(\k)$ piece in Eq.~\ref{eq:K_bar}. Note that the issue of the non-unitarity of the restricted gauge transformation is more prominent in a topological band since the quantum metric is bounded from below by the Berry curvature. Generically, for any band with a non-vanishing quantum metric, even when it is topologically trivial, the issue arises due to the fact that the gauge transformation is not diagonal in the band basis. Therefore, in order to  calculate the electromagnetic response correctly, it is important to ``integrate over'' the high energy degrees of freedom rather than simply ``projecting'' them out, as we discuss in Sec.~\ref{sec:theory} below using two different approaches.

\subsection{Transverse vs. longitudinal response in the limit of zero wavevector} 
\label{limits}
Before moving on to the effective response functions for interacting electron systems, we elaborate briefly on the difference between various limiting cases. As we demonstrated in Sec.~\ref{sec:non-int-multi} above, different ways of approaching the limit of $\q \rightarrow 0$ and $\omega \rightarrow 0$ can yield different results. To gain a better understanding of this, let us begin with the current operator,
\begin{equation}
    J_\mu(\q) = \sum_{\k,m,m'} c_{\k+\frac{\q}{2}m}^\dag c^{\phantom\dagger}_{\k-\frac{\q}{2}m'}
    \langle u_{\k+\frac{\q}{2},m} | \partial_{k_\mu} \hat{h}_\k|u_{\k-\frac{\q}{2},m'}\rangle,
    \label{eq:current_uv}
\end{equation}
and consider the non-commuting limits of $\q \rightarrow 0$ vs. $\omega \rightarrow 0$. 
In the limit of $\q \rightarrow 0$, schematically $J_\mu(\q) = J_\mu(\q = 0) + \q \cdot \left(\partial_\q J_\mu(\q)\right)_{\q=0}+ O(\q^2)$. 
 Although the difference between $\q = 0$ and $\q \rightarrow 0$ is only proportional to $\q$, we cannot set $\q = 0$ from the outset when evaluating $\cx(\q \rightarrow 0, \omega = 0)$ since the propagator can diverge as $1/q$. This is precisely why we get a finite Drude weight but a zero superfluid stiffness for multi-band non-interacting electrons in Sec.~\ref{sec:non-int-multi}.

Let us now discuss the distinction between transverse versus longitudinal current response. For the longitudinal current at long wavelength, say $J_x(q_x \rightarrow 0)$, it can be related to the commutator of $\hk$ with the density operator at finite wavelength, $\hat{\rho}_{q_x} = \sum_{\k} c_{\k,m}^\dag c_{\k - q_x \hat{e}_x ,m'}^{\phantom\dagger} \langle u_{\k,m}| u_{\k - q_x \hat{e}_x ,m'}\rangle $, where $\hat{e}_x$ is the unit vector along $x$-direction. The commutator is given by,
\begin{subequations}
 \beq
   [ \hat{\rho}_{q_x}, \hk]&=& \sum_{\k,m,m'} c_{\k , m}^\dag c_{\k - q_x\hat{e}_x , m'}^{\phantom\dagger}  \langle u_{\k,m} | u_{\k - q_x \hat{e}_x ,m'}\rangle (\varepsilon_{\k- q_x \hat{e}_x ,m'} -\varepsilon_{\k ,m})\\
   &=&\sum_{\k,m,m'} c_{\k , m}^\dag c_{\k - q_x\hat{e}_x , m'}^{\phantom\dagger}  \langle u_{\k,m} |\hat{h}_{\k- q_x \hat{e}_x} - \hat{h}_{\k} | u_{\k - q_x \hat{e}_x ,m'}\rangle\\
   &=&\sum_{\k,m,m'} c_{\k+ \frac{q_x \hat{e}_x}{2}, m}^\dag c_{\k - \frac{q_x \hat{e}_x}{2} , m'}^{\phantom\dagger}  \langle u_{\k+\frac{q_x \hat{e}_x}{2},m} |\hat{h}_{\k- \frac{q_x \hat{e}_x}{2}} - \hat{h}_{\k + \frac{q_x \hat{e}_x}{2}} | u_{\k - \frac{q_x \hat{e}_x}{2} ,m'}\rangle. \nn\\
\eeq   
\end{subequations}
Expanding around $q_x =0$, we obtain
 \beq
[ \hat{\rho}_{q_x}, \hk]&=& -q_x \sum_{\k,m,m'} c_{\k+ \frac{q_x \hat{e}_x}{2}, m}^\dag c_{\k - \frac{q_x \hat{e}_x}{2} , m'}^{\phantom\dagger}  \langle u_{\k+\frac{q_x \hat{e}_x}{2},m} |\partial_{k_x}\hat{h}_{\k} | u_{\k - \frac{q_x \hat{e}_x}{2} ,m'}\rangle + O(q_x^2)\nn\\
& =& -q_x J_x(q_x \hat{e}_x) + O(q_x^2).
 \eeq
We can therefore write $J_x(q_x \rightarrow 0)$ as,
\beq
J_x(q_x \rightarrow 0) = - \lim_{q_x \rightarrow 0 }\frac{d [ \hat{\rho}_{q_x}, \hk]}{d q_x} \equiv - i [\hat{X}, \hk],
\label{eq:current_transverse}
\eeq
since the matrix element of $\hat{\rho}_{q_x}$ between two states is ultimately $\langle m| e^{i q_x X} |n\rangle$, where $\hat{X}$ is the many-body position operator as before. 

From Eq.~\ref{eq:current_transverse}, it is readily seen that the longitudinal response should vanish by considering an infinitesimal unitary transformation $U_\alpha  = e^{i \alpha \hat{X}}$. Let us denote the Hamiltonian and its ground state after applying $U_\alpha$ as $H[\alpha]$ and $| \psi_\alpha\rangle$, respectively. The longitudinal response is then proportional to $\partial_\alpha^2 \left(\langle \psi_\alpha | H[\alpha] | \psi_{\alpha}\rangle\right)$, which vanishes following the unitarity of $U_\alpha$.

On the other hand, the transverse current, say $J_x(q_y \hat{e}_y)$, cannot directly be related to the commutator of the Hamiltonian with the many-body position operator. Therefore we need to refer to the original definition in Eq.~\ref{eq:current_uv} and keep track of the non-commuting limits when calculating $\chi_{xx}(q_y \rightarrow 0, \omega =0)$, as we highlight below in the general interacting problem.

\section{Theoretical results} 
\label{sec:theory}

Given a band structure of the form shown in Fig.~\ref{fig:many_TBG_bands}(a), we are interested in the low-energy electromagnetic response of partially filled bands (upto an energy $E_F$), with a characteristic bandwidth $W$, separated from the remote bands by an energy-gap, $\Delta$. As noted in our earlier discussion in the context of twisted bilayer graphene in Fig.~\ref{fig:many_TBG_bands}, the ``active'' bands could be any subset of the narrow and dispersive bands.
We then proceed to decompose the many-body Hamiltonian into two pieces, using the projection operators, $\mathbb{P}$ and $\mathbb{Q}=\mathbb{I}-\mathbb{P}$,
\begin{subequations}
\beq
H &=& H_d + H_o, \label{projection}\\
H_d &=& \mathbb{P} H \mathbb{P} + \mathbb{Q} H \mathbb{Q},\\
H_o &=& \mathbb{P} H \mathbb{Q} + \mathbb{Q} H \mathbb{P}. 
\eeq
\end{subequations}
The operator $\mathbb{P}$ projects the many-body states to the Hilbert space, $\mathbb{H}$, with partially occupied active bands and fully occupied (empty) lower energy (higher energy) remote bands; see Fig.~\ref{fig:many_TBG_bands}(a). The action of the different terms in the many-body Hamiltonian is shown schematically in Fig.~\ref{fig:many_TBG_bands}(a). 
In the absence of the external probe gauge field, the effective Hamiltonian is given by $\heff \equiv \mathbb{P} H\mathbb{P} = \mathbb{P} H_d\mathbb{P}$. However, coupling the theory to an external vector potential induces ``mixing'' between the different sectors of the theory. The next two subsections provide a complementary understanding of organizing the leading non-trivial contributions of such mixing terms to the diamagnetic response. 

\subsection{Review of Schrieffer-Wolff and projected gauge transformations}
\label{sec:SW}

In a recent article \cite{DMDC1}, we developed a formalism to correctly evaluate $\langle\kxx\rangle$ using a Schrieffer-Wolff (SW) type transformation \cite{SW_rev} in the presence of a small probe gauge-field, $\vec{A}$. In Sec.~\ref{sec:pert}, we will present a complementary method that arrives at the same set of results. For pedagogical reasons and to make the contrast between the two methods clear, it is useful to provide a brief summary of the SW transformation here; see Ref.~\cite{DMDC1} for further details. We express $H[A] \equiv H_d[A] + H_o[A]$, where $H_o[A]$ couples together the active and remote bands. Since the goal is to obtain an effective Hamiltonian that does not have matrix elements between these two sets of bands, we carry out a SW transformation, $\widetilde{H}[A] = e^{ T[A]} H[A] e^{-T[A]}$, where $\langle m | T[A] |n \rangle = \langle m |H_o[A] |n\rangle/(E_m - E_n)$. Here $m,n$ correspond to energy levels of $H_d[A]$ with $|E_m - E_n| \geq \Delta$. The resulting $\heff[A]\equiv \mathbb{P}\widetilde{H}[A]\mathbb{P}$, introduced in Eq.~\ref{Ds}, is given by:
\beq
\heff[A] =  \mathbb{P}H_d[A] \mathbb{P}    
 +\frac{1}{2}  \sum_{\substack{m,n\in\mathbb{H},\\ \l\notin \mathbb{H}}} \bigg[\langle m |H_o[A] |\l\rangle \langle \l| H_o[A] |n\rangle \times \left(\frac{1}{E_m -E_\l} - \frac{1}{E_\l - E_n}\right)\bigg]  + ...,\nn\\\label{eq:H_eff_SW}
\eeq
where we ignore higher order terms in $1/\Delta$ and $A$. The effective current, $\J_\tn{x}$, susceptibility, $\cxx$, and diamagnetic contribution, $\kxx$, can be obtained by expanding Eq.~\ref{eq:H_eff_SW} up to second order in $A$ and calculating the appropriate derivatives as in Eq.\ref{susc}. In the absence of an external $A$, note that ($\mathbb{P}H\mathbb{P}$) is independent of $\Delta$, while the second term in Eq.~\ref{eq:H_eff_SW} above is  $O(V^2/\Delta)$. For $A\neq 0$, both of the terms above contribute an $O(\Delta)$ correction to the $O(A^2)$ terms. As shown in Ref.~\cite{DMDC1}, these corrections have to cancel out as the effective low-energy theory should be perfectly well defined in the $\Delta\rightarrow\infty$ limit.

The effective current operator and the diamagnetic response can be obtained by taking an appropriate number of derivatives with respect to $A$, leading to
\beq
    \J_\mu(\q \rightarrow 0) &=& \mathbb{P} \left(J_\mu (\q\rightarrow 0)  + i \left[\hat{X}_\mu, H_o\right]\right) \mathbb{P}
    \label{eq:J_eff},\\
    \langle \kxx \rangle &=& - \bigg\langle\left[\mathbb{P}\hat{X}\mathbb{P}, \left[\mathbb{P}\hat{X}\mathbb{P}, \mathbb{P} H_d \mathbb{P}\right]\right]\bigg\rangle.
   \label{eq:K_xx_gauge}
\eeq
For the effective current operator, the second term arises from mixing between active and remote bands \cite{sondhi1992long}. Moreover, the above results indicate that ultimately only the projected degrees of freedom enter both quantities. As was emphasized in Ref.~\cite{DMDC1}, we can interpret Eq.~\ref{eq:K_xx_gauge} as the response arising from an ``effective'' gauge transformation, $U_\alpha^{\tn{eff}} = e^{i \alpha \mathbb{P} \hat{X}\mathbb{P}}$, involving the projected position operator, $\mathbb{P}\hat{X}\mathbb{P}$. The projected gauge-transformation is natural, given that there is an emergent conservation law associated with only the number of electrons in the flat-bands.

In contrast, the ``naive'' analysis for computing the diamagnetic response would have been the following: project interactions to the flat-bands (in the absence of a vector-potential) to obtain the effective theory and then carry out a simple gauge-transformation for all the electrons, where the vector-potential couples to the global-charge, via
\beq
\langle \kxxnaive \rangle \equiv \lim_{\alpha \rightarrow 0} \partial_\alpha^2 \langle U_\alpha^{\rm{naive}}~\mathbb{P} H_d \mathbb{P}~(U_\alpha^{\rm{naive}})^\dagger \rangle,
\label{eq:Kxxnaive}
\eeq
where $U_\alpha^{\rm{naive}} = e^{i \alpha \hat{X}}$ is the standard gauge-transformation associated with minimal-coupling. Note that $\langle \kxxnaive \rangle$ cannot be viewed in general as a ``looser'' bound on the superfluid stiffness, since its magnitude is not necessarily larger than $\langle \kxx \rangle$.

In the next subsection, we will arrive at the same set of results using a different method.

\subsection{Low-energy electromagnetic response from perturbation theory}
\label{sec:pert}

In this section, we present an alternative approach, which is best thought of as a  perturbative expansion in $(V/\Delta)$;  however, in what follows we will also be able to obtain the important $O(\Delta)$ terms explicitly. Let us denote the ground state of the full Hamiltonian, $H$, as $H \gs = E_0 \gs $. Clearly, the ground state of the effective Hamiltonian $\heff \equiv \mathbb{P} H \mathbb{P} = \mathbb{P} H_d \mathbb{P}$ is also the ground state of $H_d$, i.e. $H_d \g0 = \ef \g0 $. We can then obtain, 
\beq
    \gs &=& \g0 - \mathbb{Q}\frac{1}{H_d - \ef}\mathbb{Q}H_o\mathbb{P}\g0 + O\left[(V/\Delta)^2\right],\\
    E_0 &=& \ef -  \langle \tn{GS}_0 |\mathbb{P}H_o\mathbb{Q} \frac{1}{H_d - \ef}\mathbb{Q}H_o\mathbb{P}\g0 .
\eeq 
Analogously, we can also perform a perturbative expansion of the response function. 
Let us begin with the expression for the full diamagnetic response, where starting with Eq.~\ref{kxx} we obtain,
\beq
\langle K_{\tn{xx}} \rangle = 2 \langle \tn{GS} | \hat{X} (H - E_0) \hat{X} \gs.
\eeq

Similarly, the full expression for the paramagnetic current susceptibility, $\chi_{\tn{xx}}$, can be obtained using the Lehmann representation and setting $\omega=0$, 
\beq
        \chi_{\tn{xx}}(q\rightarrow 0, \omega =0) =2 \lim_{q\rightarrow 0 } \langle \tn{GS} |J_\tn{x}(q) \frac{1}{H - E_0} J_\tn{x}(-q) \gs.
\eeq
Note that different ways of approaching $q\rightarrow 0$ can yield different values of $\chi_{\tn{xx}}$, and is controlled by the behavior of the propagator $(H- E_0)^{-1}$ around its pole, which only matters when $J_\tn{x}(q)$ acts within the active bands. This can already be seen based on the discussion in Secs. \ref{sec:non-int-multi} and \ref{limits}. Hence it is convenient to express $J_\tn{x}(q)$ in terms of its projection $J_\tn{x}^0$, and other ``inter-band'' contribution, $J_\tn{x}^\tn{int}$ as,
\beq
\lim_{q\rightarrow 0 } J_\tn{x}(q) &=& \lim_{q\rightarrow 0 } \mathbb{P} J_\tn{x}(q) \mathbb{P} - i\mathbb{P} [\hat{X},H] \mathbb{Q}  - i\mathbb{Q} [\hat{X},H] \mathbb{P} - i\mathbb{Q} [\hat{X},H] \mathbb{Q} \\
&=& \lim_{q\rightarrow 0 } \mathbb{P} J_\tn{x}(q) \mathbb{P} - i [\hat{X},H]  + i\mathbb{P} [\hat{X},H] \mathbb{P}\\
&\equiv& J_\tn{x}^0 + J_\tn{x}^\tn{int},
\eeq
where $J_\tn{x}^0 = \lim_{q\rightarrow 0 } \mathbb{P} J_\tn{x}(q) \mathbb{P}$.

Let us first consider the contribution from $J_\tn{x}^\tn{int}$,
\begin{subequations}
\beq
&&\langle \tn{GS}| J_\tn{x}^\tn{int} (H-E_0)^{-1} J_\tn{x}^\tn{int} \gs \nn\\
&=& - \langle \tn{GS}| [\hat{X}, H] (H-E_0)^{-1}  [\hat{X}, H] \gs  \nn\\&& + 2 \tn{Re}\left[\langle \tn{GS}|  [\hat{X}, H]  (H-E_0)^{-1}\mathbb{P}  [\hat{X}, H]\mathbb{P} \gs \right] \nn\\ &&- \langle \tn{GS}| \mathbb{P} [\hat{X}, H]\mathbb{P}  (H-E_0)^{-1}\mathbb{P}  [\hat{X}, H]\mathbb{P} \gs
\\&=& \langle \tn{GS}| \hat{X} (H - E_0) \hat{X} \gs + 2 \tn{Re}\left[\langle \tn{GS}|  \hat{X} \mathbb{P}  [\hat{X}, H]\mathbb{P} \gs \right] \nn\\
&& - \langle \tn{GS}| \mathbb{P} [\hat{X}, H]\mathbb{P}  (H-E_0)^{-1}\mathbb{P}  [\hat{X}, H]\mathbb{P} \gs.
\eeq
\end{subequations}

Note that the first term above is exactly proportional to $\langle K_{\tn{xx}} \rangle$. 
Since $\mathbb{P} [\hat{X}, H]\mathbb{P} \sim O(V)$, we only need to retain terms that are independent of $\Delta$ in $\mathbb{P}\gs $ and $\mathbb{P} (H-E_0)^{-1} \mathbb{P}$, which are $\g0 $ and $\mathbb{P} (H_d-\ef)^{-1} \mathbb{P}$, respectively. Therefore,
\begin{subequations}
\beq
&&2 \langle \tn{GS}| J_\tn{x}^\tn{int} (H-E_0)^{-1} J_\tn{x}^\tn{int} \gs - \langle K_{\tn{xx}} \rangle \nn\\
&=&  -4\langle \tn{GS}_0 |  \hat{X} \mathbb{P}(H_d - \ef)\mathbb{P}  \hat{X} \g0  +  4 \tn{Re}\left[\langle \tn{GS}_0 |  \hat{X} \mathbb{P}  [\hat{X}, H_o]\mathbb{P} \g0 \right]\nn\\
&& - 2\langle \tn{GS}_0 | \mathbb{P} [\hat{X}, H_d + H_o]\mathbb{P}  (H_d-\ef)^{-1}\mathbb{P}  [\hat{X}, H_d + H_o]\mathbb{P} \g0 \\
&=& - 2 \langle \tn{GS}_0 |  \hat{X} \mathbb{P}(H_d - \ef)\mathbb{P}  \hat{X} \g0  - 2 \langle \tn{GS}_0 |  [\hat{X},  H_o]\mathbb{P}  \frac{1}{ H_d-\ef}\mathbb{P}  [\hat{X}, H_o] \g0 \nn\\
&&~~~~~~~~~~~~~~~~~~~~~~~~~~~~~~~~~~~~~~~~~~~~~~~~~~~~~~~~~~~~~~~~~~+ O\bigg(\frac{V^2}{\Delta}\bigg).
\eeq
\end{subequations}

Next, we consider the mixing term between $J_\tn{x}^\tn{int}$ and $J_\tn{x}^0$,
\begin{subequations}
\beq
&&\langle \tn{GS}| J_\tn{x}^0 (H-E_0)^{-1} J_\tn{x}^\tn{int} \gs \nn\\
&=& -i  \langle \tn{GS}| J_\tn{x}^0 \mathbb{P} (H-E_0)^{-1} [\hat{X}, H]  \gs + i  \langle \tn{GS}| J_\tn{x}^0 \mathbb{P} (H-E_0)^{-1} \mathbb{P}[\hat{X}, H]\mathbb{P} \gs \\
&=&  i  \langle \tn{GS}| J_\tn{x}^0 \mathbb{P} \hat{X}  \gs + i \langle \tn{GS}_0| J_\tn{x}^0 \mathbb{P}(H_d-\ef)^{-1} \mathbb{P}[\hat{X}, H_d + H_o]\mathbb{P}\g0  \\
&=& i \langle \tn{GS}_0| J_\tn{x}^0 \mathbb{P}(H_d-\ef )^{-1} \mathbb{P}[\hat{X}, H_o]\mathbb{P}\g0 + O\bigg(\frac{V^2}{\Delta}\bigg). 
\eeq
\end{subequations}

Finally, we combine all the terms together to obtain
\begin{subequations}
\beq
&&\langle K_{\tn{xx}} \rangle - \chi_{\tn{xx}}(q\rightarrow 0, \omega =0) = \langle K_{\tn{xx}} \rangle - 2 \langle \tn{GS}| (J_\tn{x}^0+ J_\tn{x}^\tn{int}) (H-E_0)^{-1} (J_\tn{x}^0+ J_\tn{x}^\tn{int}) \gs \nn \\ \\
&=&  2 \langle \tn{GS}_0 |  \hat{X} \mathbb{P}(H_d - \ef)\mathbb{P}  \hat{X} \g0  + 2 \langle \tn{GS}_0 |  [\hat{X},  H_o]\mathbb{P}  \frac{1}{ H_d-\ef}\mathbb{P}  [\hat{X}, H_o] \g0 \nn\\
&&+ 2 \tn{Re}\left[- i \langle \tn{GS}_0| J_\tn{x}^0 \mathbb{P}(H_d-\ef )^{-1} \mathbb{P}[\hat{X}, H_o]\mathbb{P}\g0 \right]\nn\\
&&- 2 \langle \tn{GS}_0 J_\tn{x}^0 (H_d-\ef )^{-1} J_\tn{x}^0 \g0 \\
&=&  2 \langle \tn{GS}_0 |  \hat{X} \mathbb{P}(H_d - \ef)\mathbb{P}  \hat{X} \g0 \nn\\&& - 2 \lim_{q\rightarrow 0 }\langle \tn{GS}_0|  (\mathbb{P} J_\tn{x}(q) \mathbb{P} + i [\hat{X}, H_o]) (H_d-\ef )^{-1} (\mathbb{P} J_\tn{x}(q) \mathbb{P} + i [\hat{X}, H_o]) \g0 + O\bigg(\frac{V^2}{\Delta}\bigg),\nn\\
\eeq
\end{subequations}
which agrees with our previous result obtained using the SW transformation in Sec.~\ref{sec:SW}; see Eq.~\ref{eq:J_eff} and \ref{eq:K_xx_gauge}. Although being conceptually the same, there is a practical difference between the two approaches. In the perturbative approach of computing the response functions as discussed in this section, the effective current operator is constructed retrospectively, whose form is unknown {\it a priori}. On the other hand, the Schrieffer-Wolff transformation naturally yields the effective current operator directly.

\section{Approximate upper bounds on superconducting $T_c$}
\label{sec:bounds}

The formalism we have developed in Sec.~\ref{sec:theory} is useful for clarifying a number of issues with regards to computing the electromagnetic response of generic interacting models. We focus now on a practical application of this formalism to evaluate the maximum possible superconducting $T_c$. The remainder of this section is organized as follows: In the next few subsections, we apply the formalism to evaluate (approximate) upper bounds on $T_c$ for various sign-problem free models where the actual $T_c$ is already known from unbiased QMC simulations.   

\begin{figure}
    \centering
    \includegraphics[width =0.9  \textwidth]{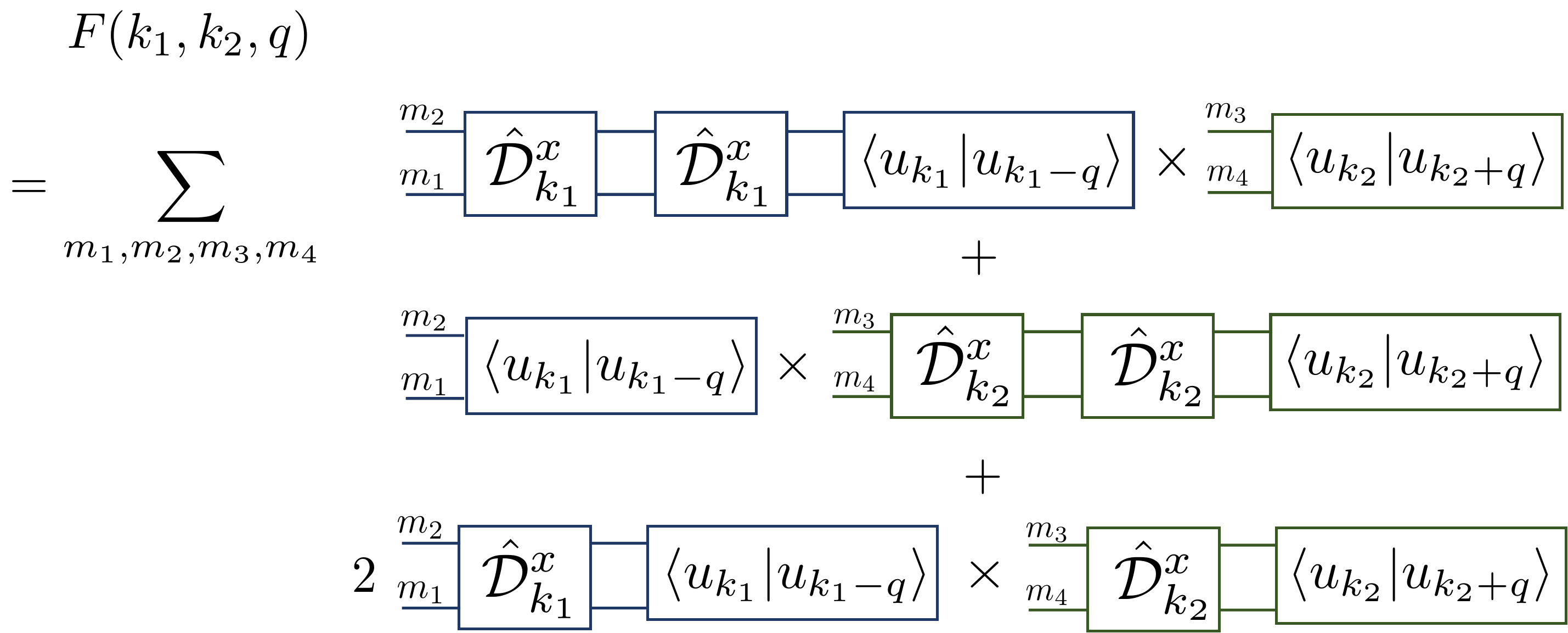}
    \caption{A schematic expression for $F(\k_1,\k_2,\q)$ in Eq.~\ref{F}. The labels $m_i$ represent orbital indices and the solid lines that link the blocks denote the different index contraction.}
    \label{fig:fk1k2q}
\end{figure}

As noted above, in two-dimensions $T_c = \pi D_s(T_c^- )/2$. Since the effective paramagnetic term $\cxx$ is positive semi-definite, $D_s(T) \leq \frac{e^2}{4} \langle \kxx \rangle_T$, where  $\langle ... \rangle_T$ denotes the thermal average in the many-body state. Importantly, both the kinetic energy and the interaction terms contribute to $\langle \kxx \rangle_T$, which involves evaluating a multi-particle expectation value. Doing this exactly for a given Hamiltonian is typically impossible, unless a detailed knowledge of the actual many-body state is already available. The general expression for Hamiltonians of the form in Eq.\ref{ham1} can be written as,
\beq
\langle\kxx\rangle_T = \sum_{\k,m,m'} \frac{\partial^2\widetilde{\ve}_{\k m m'}}{\partial k_x^2} \langle c^\dagger_{\k m} c^{\phantom\dagger}_{\k m'}\rangle + \sum_{\k_1,\k_2,\q} V(\q) F(\k_1,\k_2,\q) \langle c_{\k_1m}^\dag c_{\k_2m'}^\dag c^{\phantom\dagger}_{\k_2+\q m''} c^{\phantom\dagger}_{\k_1-\q m'''} \rangle,\label{kxx1}
\eeq
where
\beq
\widetilde{\ve}_{\k m m'} = \ve_{\k m} \delta_{mm'} + \sum_{\q,n} V(\q) \langle u_{\k m}|u_{\k-\q n}\rangle \langle u_{\k-\q n}| u_{\k m'}\rangle,
\eeq
is the electronic dispersion including the renormalization from the interaction and
\beq
F(\k_1,\k_2,\q) =  \left[ \hat{\mathcal{D}}_{\k_1}^x + \hat{\mathcal{D}}_{\k_2}^x\right]^2 \langle u_{\k_1 m} | u_{\k_1-\q m'''}
\rangle \langle u_{\k_2 m'}| u_{\k_2 +\q m''}\rangle.
\label{F}
\eeq
Note that both of the interaction-induced contributions arise from normal-ordering. Here $\hat{\mathcal{D}}_{\k_1}^\mu$ is a 4-leg tensor operator acting on the form factor (see Fig. \ref{fig:fk1k2q} for an illustration), which can be viewed as a ``covariant'' derivative in $\k-$space and  $\hat{\mathcal{D}}_{\k_1}^\mu$ acting on the form factor as,
\beq
\left[\hat{\mathcal{D}}_{\k}^\mu\right]_{mm',nn'} \langle u_{\k,m'} | u_{\k-\q,n'} \rangle  = (\partial_{k_\mu} \delta_{mm'} \delta_{nn'}- i \mathcal{A}_{\k,mm'}^\mu \delta_{nn'} + i \mathcal{A}_{\k-\q,n'n}^\mu \delta_{mm'} ) \langle u_{\k,m'} | u_{\k-\q,n'}\rangle, \nn\\
\label{eq:cov_D_k}
\eeq
with $\mathcal{A}_{\k,mm'}^\mu = i \langle u_{\k,m} | \partial_{k_\mu} u_{\k,m'}\rangle$ the multi-orbital Berry connection.

Since we are only interested in obtaining an upper bound, we can replace the many-body expectation values by a constant that depends only on the band filling. We can bound the expectation value $\langle c_\alpha^\dagger c_\beta^\dagger c^{\phantom\dagger}_\gamma c^{\phantom\dagger}_\delta \rangle \leq C(\nu)\equiv  \tn{min}\{\nu, n_{\tn{max}}-\nu\}$, that depends only on the filling $\nu$ relative to the maximum filling, $n_{\tn{max}}$, associated with the active-band \cite{MR21}; the indices denote any combination of momentum and orbital labels. Note that for the sake of simplicity we are effectively evaluating these quantities at $T=0$ with the assumption that $D_s(0)$ provides an approximate upper bound on $T_c$; this is not necessarily guaranteed for all microscopic models \cite{heuristicbound}. However, for the microscopic models where we apply our framework, we compare our estimated upper bound on $D_s(0)$ with the same quantity obtained directly from QMC. 

Finally, we obtain,
\beq
D_s\leq C(\nu) \left(\frac14  \sum_{\k} \left|\frac{\partial^2 \widetilde{\varepsilon}_{\k }}{\partial k_x^2}\right| + \frac14 \sum_{\k_1,\k_2,\q}| V(\q)F(\k_1,\k_2,\q)|\right) (\equiv D_s^{\tn{upper}}).
\label{dsupper}
\eeq

\subsection{Sign-problem free model with topological Chern bands}
\label{sec:qmc}

\begin{figure}
    \centering
    \includegraphics[width=0.95\textwidth]{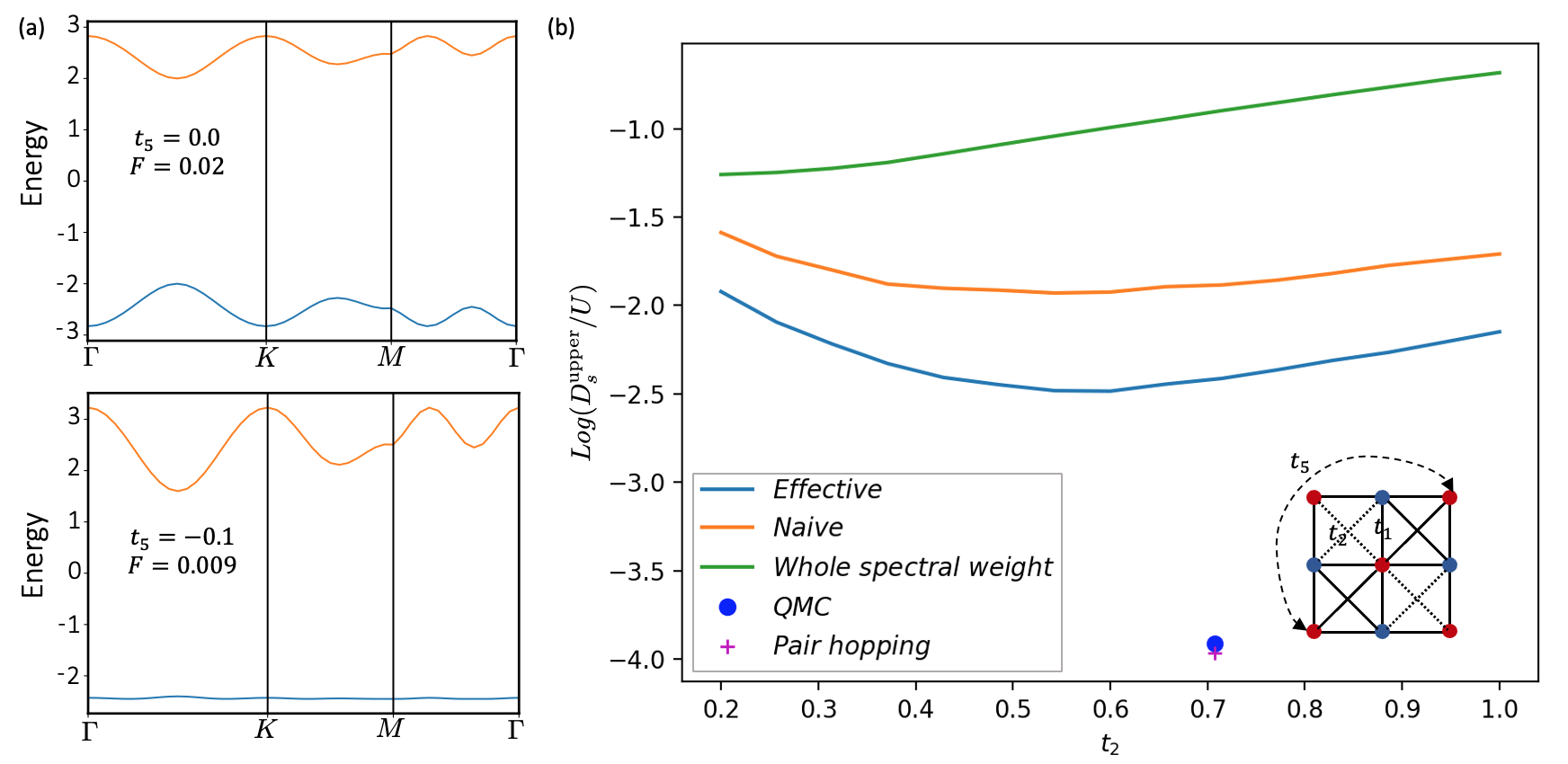}
    \caption{(a) Electronic dispersion for the lattice model in Eq.~\ref{piflux} for two different values of $t_5$ with $t_1=1$ and $t_2 = 1/\sqrt{2}$. $F=W/\Delta_{\rm{gap}}$ denotes the flatness-ratio for the lower band. (b) The numerically evaluated value for $D_s^{\rm{upper}}/U$ obtained from Eq.~\ref{dsupper} (on a log-scale) as a function of $t_2$. The inset depicts the hopping parameters associated with the model in Eq.~\ref{piflux}; solid (dashed) lines between second nearest neighbor denotes the sign structure. The blue, orange and green curves denote the upper bound obtained using the ``effective'' gauge transformation (Eq.~\ref{eq:K_xx_gauge}), the ``naive'' gauge transformation (Eq.~\ref{eq:Kxxnaive}) and contribution from the full spectral weight involving both bands (for $U = t_1$). The blue dot denotes the QMC result from Ref.~\cite{hofmann2020superconductivity} and the magenta cross denotes the contribution from {\it only} the pair-hopping terms after performing a wannierization for the flat-bands (see Ref.~\cite{MR21} for details).} 
    \label{fig:pi-flux-model}
\end{figure}

The first unambiguous (and unbiased) demonstration of superconductivity in a flat-band system was presented in Ref.~\cite{hofmann2020superconductivity} in a model of topological bands with an on-site attractive Hubbard interaction. The time-reversal symmetric model consists of $C=\pm1$ bands, where the different Chern bands are tied to opposite spins, $\sigma=\uparrow,~\downarrow$. The explicit lattice model, $H = H_{\tn{kin}} + H_{\tn{int}}$, defined on a 2D square lattice is given by ($U>0$),
\begin{subequations}
\beq
H_{\tn{kin}} &=& \bigg[-t_1\sum_{\la i,j\ra,\sigma} e^{i\phi^\sigma_{ij}} c^\dag_{i,\sigma} c^{\phantom{\dag}}_{j,\sigma}  - t_2 \sum_{\la i,j\ra_2,\sigma} s_{\la i,j\ra_2} c^\dag_{i,\sigma} c^{\phantom{\dag}}_{j,\sigma}  - t_5 \sum_{\la i,j\ra_5,\sigma} c^\dag_{i,\sigma} c^{\phantom{\dag}}_{j,\sigma} + \tn{h.c.}\bigg] -\mu\sum_i n_i  \label{piflux}\nn \\ \\
H_{\tn{int}} &=& -\frac{U}{2}\sum_i (n_i - 1)^2,
\eeq
\end{subequations}
where, $c^\dagger_{i,\sigma}$ ($c_{i,\sigma}$) are fermion creation (annihilation) operators, $n_i=\sum_\sigma c^\dagger_{i\sigma}c_{i\sigma}$ is the local density, $t_1,~t_2,~t_5$ denote first, second and fifth nearest neighbor hopping parameters, respectively (see inset of Fig.~\ref{fig:pi-flux-model}b). The second-neighbor hopping, $t_2$, differs in sign between unit cells and  $\phi_{ij}^\sigma = -\phi_{ij}^{-\sigma}$, with $\phi_{ij}^\uar{} = \pm \frac{\pi}{4}$. The density is tuned by varying the chemical potential, $\mu$. In the numerical study, results for various correlation functions (including $D_s$) were obtained for two different bare bandwidths for the flat-bands, with ``flatness-ratios'', $F \equiv W/\Delta_{\tn{gap}}= 0.2,~0.009$; see Fig.\ref{fig:pi-flux-model} (a).

We now obtain a conservative upper bound on the value of $D_s\leq  \frac{e^2}{4} \langle \kxx\rangle$, at $T=0$. While $T_c$ is only determined by $D_s(T_c^-)$, our bounds are approximate enough that these subtleties will not affect the results. We obtain these approximate bounds starting with (i) the full theory, (ii) the projected model based on the ``correct'' (i.e. effective) gauge transformation, and (iii) the projected model based on the ``incorrect'' (i.e. naive) gauge transformation. Clearly, the bound based on (i) is not particularly useful since it includes the full optical spectral weight (including contribution from the higher band and inter-band matrix elements); see Fig.~\ref{fig:pi-flux-model}(b). 

We plot $D_s^{\tn{upper}}$ in Fig.~\ref{fig:pi-flux-model}(b) as a function of $t_2$. The main effect of varying $t_2$ at the level of the non-interacting band structure is to change the quantum metric and Berry curvature distribution in the Brillouin zone. Although not directly related to $D_s$, the integrated quantum metric has a similar trend to the bound $D_s^{\tn{upper}}$. Relatedly, the momentum sum in Eq.~\ref{dsupper} includes contributions from terms that are not simply determined by the quantum metric. Interestingly, the extremely conservative and approximate bound we obtain is about four times the exact value for $D_s(T=0)$ obtained from the QMC computations (which is $\approx 0.02 U$). The numerical integrals were performed using the vegas package in Python \cite{vegas_2022}.

\subsection{Sign-problem free model with fragile topological bands}
Going beyond topological Chern bands, a recent QMC based analysis was also employed to study superconductivity in flat-bands with a ``fragile'' topology and an on-site attractive Hubbard interaction \cite{peri2021fragile}. The Hamiltonian defined on a kagome lattice is of the form,
\begin{subequations}
\beq
\hk &=& \sum_{\k,\sigma} c_{\k,\sigma}^\dag \hat{h}_\k c_{\k,\sigma},\\
\hi &=& - \frac{U}{2} \sum_i (n_i - 1)^2,
\eeq
\end{subequations}
where $n_i = \sum_\sigma c_{i \sigma}^\dag c_{i \sigma}$ is the local density and the spin-up and spin-down electrons have same kinetic Hamiltonian, $\hat{h}_\k$, defined in momentum-space. 
\begin{figure}
    \centering
    \includegraphics[width = 0.45\textwidth]{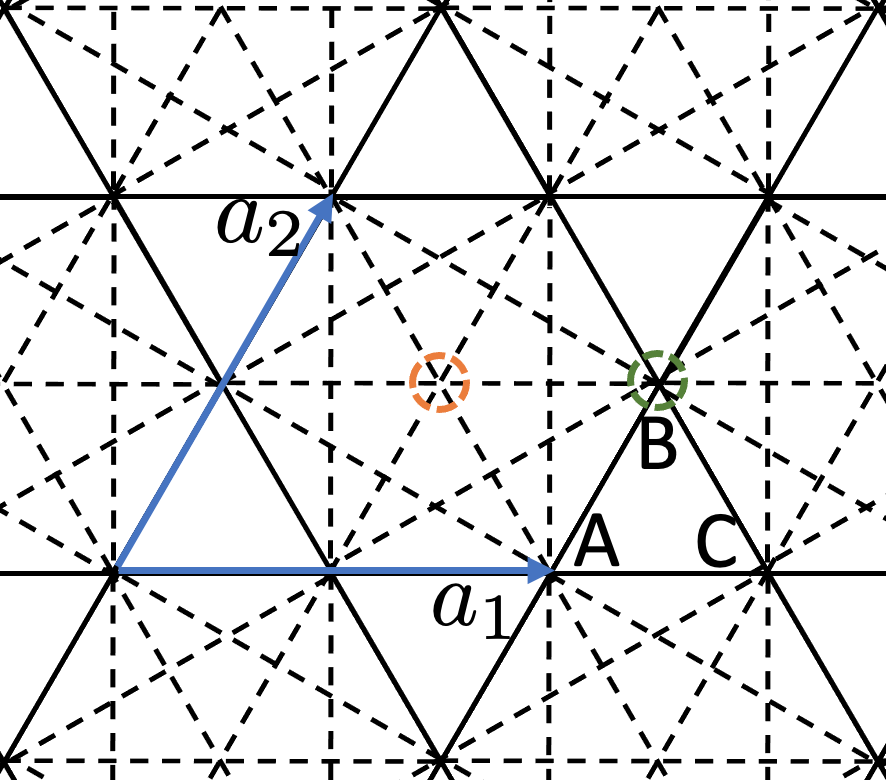}
    \caption{Kagome lattice model with three sites per unit cell, labelled A, B and C, and unit vectors, $\vec{a}_1$ and $\vec{a}_2$, respectively. The ``kagome-3'' model includes nearest neighbor hoppings and further range hoppings indicated by dashed lines, respectively; see Eq.~\ref{k3} and \cite{peri2021fragile}. The orange and green dashed circles denote the 1a and 1b Wyckoff positions, respectively.}
    \label{fig:kagome}
\end{figure}
For the kagome-3 model (see \cite{peri2021fragile} and references therein),
\beq
\hat{h}_\k =
\begin{pmatrix}
    2 \cos{\k \cdot \vec{a}_2} & f_1(\k)  & f_2(\k)\\
    f_1^*(\k) & 2 \cos{\k \cdot (\vec{a}_1 - \vec{a}_2)} & f_3(\k)\\
    f_2^*(\k) & f_3^*(\k) & 2 \cos{\k \cdot \vec{a}_1}
\end{pmatrix}
\label{k3}
\eeq
where $f_1(\k) = 1+ e^{i \k \cdot \vec{a}_1} + e^{i \k \cdot \vec{a}_2} + e^{i \k \cdot (\vec{a}_1- \vec{a}_2)}$, $f_2(\k) = 1+ e^{i \k \cdot \vec{a}_1} + e^{-i \k \cdot \vec{a}_2} + e^{i \k \cdot (\vec{a}_1- \vec{a}_2)}$ and $f_3(\k) = 1+ e^{-i \k \cdot \vec{a}_1} + e^{-i \k \cdot \vec{a}_2} + e^{i \k \cdot (\vec{a}_1- \vec{a}_2)}$, with $\vec{a}_1 = (1,0),~\vec{a}_2 = (1/2,\sqrt{3}/2)$; see Fig.~\ref{fig:kagome}. The spectrum for $\hk$ has two exactly flat bands at energy $\epsilon_\k = -2$ and a third dispersive band with a finite energy gap, $\epsilon_\k = 4 + 2 [\cos{\k \cdot \vec{a}_1} + \cos{\k \cdot \vec{a}_2} + \cos{\k \cdot (\vec{a}_1-\vec{a}_2)}]$. The two degenerate flat bands are fragile topological  \cite{peri2021fragile}. 

Additionally, by introducing a fourth orbital to the kagome-3 model and turning on a nearest neighbor hopping to the original sites, the low energy bands can be either made topologically trivial, or can remain topologically fragile depending on the position of the additional sites \cite{peri2021fragile}. If the additional sites are at the 1a Wyckoff position (Fig.~\ref{fig:kagome}), the low-energy bands become topologically trivial with three exactly flat bands at $\epsilon_\k = -2$. To be explicit,
\beq
\hat{h}_{\k,4,\tn{trivial}} = 
\begin{pmatrix}
    t^2 -2 & \vline &t (1+ e^{i \k \cdot \vec{a}_2}) & t ( e^{i \k \cdot \vec{a}_1}+ e^{i \k \cdot \vec{a}_2}) & t (1+ e^{i \k \cdot \vec{a}_1})\\
    \hline
     t (1+ e^{-i \k \cdot \vec{a}_2}) &  \vline &  &  & \\
   t ( e^{-i \k \cdot \vec{a}_1}+ e^{-i \k \cdot \vec{a}_2}) & \vline &  &\hat{h}_\k   & \\
   t (1+ e^{-i \k \cdot \vec{a}_1}) &  \vline && & 
\end{pmatrix},
\eeq
where $t$ is a hopping parameter. On the other hand, if the additional sites are at the 1b Wyckoff position, the fragile topology associated with the low-energy bands survive and there are two exactly flat bands with $\epsilon_\k = -2$ plus an additional ``nearly'' flat band. The explicit Hamiltonian is,
\beq
\hat{h}_{\k,4,\tn{fragile}} = 
\begin{pmatrix}
    m &  \vline &0 & 0&s\\
    \hline
     0 &   \vline & &  & \\
  0   &  \vline & &\hat{h}_\k   &  \\
   s &  \vline && & 
\end{pmatrix},
\eeq
where $m$ and $s$ are hopping parameters, respectively. 

We will apply our formalism to evaluate an upper bound, $D_s^{\tn{upper}}$, for the Hubbard interaction projected to these topologically fragile and trivial flat bands below.  In Table \ref{tb:kagome}, we compare the stiffness $D_s$ obtained directly from QMC in the limit of $T\rightarrow0$ with $D_s^{\tn{upper}}$ evaluated numerically in two different ways. The latter includes $D_s^{\tn{upper}}$ evaluated using the correct, as well as naive, gauge-transformation. The observed trend for our $D_s^{\tn{upper}}$ is similar to the QMC result, in that the value obtained for the topologically trivial model is much smaller than the value obtained for bands with fragile topology. We note again that for all of the above models, the naive estimate for $D_s^{\tn{upper}}$ is always larger than the correct estimate. 

\begin{table}[h!]
\centering
\begin{tabular}{ |c||c|c|c|}
 \hline
 Model &  $D_s(T=0)$ from \cite{peri2021fragile} & Effective $D_s^{\tn{upper}}$& Naive $D_s^{\tn{upper}}$ \\
 \hline\hline
 Kagome-3 &  $0.015 ~U$ & $0.076~ U$ & $0.127~ U$ \\
 \hline
 Four-band fragile & $0.017~ U$ & $0.099~ U$ & $0.161~ U$ \\
 \hline
 Four-band trivial &  $0.001~ U$ & $0.033~ U$ & $0.079~ U$ \\
 \hline
\end{tabular}
\caption{A comparison between the actual $D_s(T=0)$ from QMC in Ref.~\cite{peri2021fragile} (for a finite-sized system) and $D_s^{\tn{upper}}$. }
\label{tb:kagome}
\end{table}

\section{Extension to exciton superfluidity}
\label{sec:exciton}

The theoretical formalism developed in Secs.~\ref{sec:EM} and \ref{sec:theory} (as well as Ref.~\cite{DMDC1}) is not restricted to addressing questions about the superconducting phase-stiffness. More generally, and with minor extensions, the formalism can be adapted to compute the phase-stiffness associated with other (continuous) broken symmetries. Specifically, in this section we focus on a different form of broken symmetry associated with exciton-condensation \cite{EJPEAM,JPE_rev}, which is experimentally relevant in interacting quantum Hall and moir\'e flat-band settings  \cite{Mak2022}, where weak-coupling treatments are not necessarily reliable. As we will discuss below, the key ingredient is related to identifying the probe ``gauge-field'' that reveals the exciton phase-stiffness, which is distinct from the physical gauge-field. In the next two subsections, we first formulate the theoretical treatment necessary for computing the exciton ``diamagnetic'' response functions and then apply the formalism to the classic example of a quantum-Hall bilayer. The formalism developed here can also be applied to other excitonic superfluid candidates, regardless of the topology of the underlying system, such as moir\'e materials \cite{ma2021strongly}.

\subsection{Phase-stiffness from effective low-energy theory}
We will consider a system with a $U(1)\times U(1)$ symmetry, associated with the conservation of particle numbers for two independent species (referred to as a ``bilayer'' henceforth). In the exciton condensate, the above symmetry is spontaneously broken down to only a $U(1)$ symmetry associated with the global conservation of the total density. One of the original $U(1)$ symmetries associated with the bilayer system, which is generated by the charge difference of the two layers, is spontaneously broken. In analogy with our setup for computing the superconducting phase-stiffness associated with projected flat-bands, we can compute the exciton phase-stiffness as,
\beq
D^{\tn{eff}}_{s,\tn{exc}} = \frac{1}{4}\bigg[\la  K^{\tn{eff}}_{\tn{xx},\tn{exc}} \ra - \chi^{\tn{eff}}_{\tn{xx},\tn{exc}}(q_x = 0, q_y \rightarrow 0 ,\omega = 0)\bigg],
\label{eq:Ds_exc}
\eeq
where $ \chi^{\tn{eff}}_{\tn{xx},\tn{exc}}$ is the appropriately defined ``current-current'' correlation function,  and $K^{\tn{eff}}_{\tn{xx},\tn{exc}}$ is the ``diamagnetic'' response. These response functions can be obtained as,
\begin{subequations}
\beq
J_\mu(q_\mu \rightarrow 0) &=&  -i \left[\overline{\hat{X}}_{\mu,\tn{exc}}, H\right], \label{jmue}\\
K_{\mu\nu,\tn{exc}}^{\tn{eff}} &=& -\bigg\langle\left[\overline{\hat{X}}_{\mu,\tn{exc}}, \left[\overline{\hat{X}}_{\nu,\tn{exc}}, \mathbb{P} H\mathbb{P} \right]\right]\bigg\rangle,~\tn{where} \label{kxxe}\\
\overline{\hat{X}}_{\mu,\tn{exc}} &\equiv&  \mathbb{P} \sum_i x_i^\mu \left( c_{i,\uparrow}^\dag c_{i,\uparrow}^{\phantom\dagger} - c_{i,\downarrow}^\dag c_{i,\downarrow}^{\phantom\dagger} \right)\mathbb{P}.
\eeq
\end{subequations}
Here we have introduced a pseudospin index to label the two layers, with electron annihilation operators $c_{i,\uparrow},~c_{i,\downarrow}$ and $\mathbb{P}$ is the projection operator to the low-energy sub Hilbert space. Importantly, $\overline{\hat{X}}_{\mu,\tn{exc}}$ is the projected many-body position operator associated with $\hat{S}_z$ that determines the effective electromagnetic response in the excitonic superfluid.

Let us now focus on a concrete Hamiltonian, to evaluate specifically the contribution $K_{\mu\nu,\tn{exc}}^{\tn{eff}}$. We choose again a purely density-density interaction, as in Eq.~\ref{ham3}. It is useful to evaluate the following commutators with the projected density operators, $\overline{\rho}_{\q, \sigma} =\sum_{\k} d_{\k,\sigma}^\dag d_{\k - \q ,\sigma}  \langle u_{\k,\sigma}| u_{\k-\q,\sigma}\rangle $,
\begin{subequations}
\beq
\left[\overline{\hat{X}}_{\mu,\tn{exc}}, \overline{\rho}_{\q, \uparrow} \right] &=& i \sum_{\k} d_{\k,\uparrow}^\dag d_{\k - \q ,\uparrow} \mathcal{D}_{k_\mu} \left[ \langle u_{\k,\uparrow}| u_{\k-\q,\uparrow}\rangle\right],\\
\left[\overline{\hat{X}}_{\mu,\tn{exc}}, \overline{\rho}_{\q, \downarrow} \right] &=& - i \sum_{\k} d_{\k,\downarrow}^\dag d_{\k - \q ,\downarrow} \mathcal{D}_{k_\mu} \left[ \langle u_{\k,\downarrow}| u_{\k-\q,\downarrow}\rangle\right],
\eeq
\end{subequations}
where $d_{\k,\sigma},~d^\dagger_{\k,\sigma}$ denote the fermion annihilation and creation operators associated with the projected degrees of freedom, respectively. As before, the ``covariant derivative'',  $\mathcal{D}_{k_\mu}$, is defined as,
\beq
 \mathcal{D}_{k_\mu} \left[ \langle u_{\k,\sigma}| u_{\k-\q,\sigma}\rangle\right] \equiv \left[ \partial_{k_\mu} + i \mathcal{A}_{k,\mu} - i \mathcal{A}_{k-q,\mu} \right] \langle u_{\k,\sigma}| u_{\k-\q,\sigma}\rangle,
\eeq
where $\mathcal{A}_{k,\mu} = - i \langle u_\k |\partial_{k_\mu} u_\k \rangle$ is the Berry connection.

\subsection{Quantum Hall bilayer: lowest Landau level theory at $\nu_{\tn{tot}}=1$}

As highlighted above, a classic setting for studying excitonic condensates are bilayer quantum Hall systems \cite{JPE_rev}. Therefore, we apply our framework to compute the exciton phase-stiffness associated with a specific quantum Hall bilayer system with total filling $\nu_{\tn{tot}} = 1$. The density-density interaction has both intra-layer and inter-layer terms, respectively,
\beq
H = \frac12\sum_\q V_{\q,\uparrow} \overline{\rho}_{\q ,\uparrow}  \overline{\rho}_{-\q ,\uparrow} + \frac12\sum_\q V_{\q,\downarrow} \overline{\rho}_{\q ,\downarrow}  \overline{\rho}_{-\q ,\downarrow}  + \frac12\sum_\q V_{\q}^\perp \left(\overline{\rho}_{\q ,\uparrow}  \overline{\rho}_{-\q ,\downarrow} + \overline{\rho}_{\q ,\downarrow}  \overline{\rho}_{-\q ,\uparrow}\right).
\eeq

In the lowest Landau-level (LLL) theory, where both layers experience the same magnetic field, the form-factors are given by, $\langle u_{\k ,\uparrow} | u_{\k -\q,\uparrow} \rangle = \langle u_{\k ,\downarrow} | u_{\k -\q,\downarrow} \rangle =  e^{-|q|^2/4 + i \q\wedge \k /2}$, where $\q\wedge \k \equiv q_x k_y - q_y k_x$. Therefore, 
\beq
 \mathcal{D}_{k_x} \left[ \langle u_{\k,\uparrow}| u_{\k-\q,\uparrow}\rangle\right] = \mathcal{D}_{k_x} \left[ \langle u_{\k,\downarrow}| u_{\k-\q,\downarrow}\rangle\right] = -i q_y  e^{-|q|^2/4 + i \q\wedge \k /2}.
\eeq
It is readily seen that only terms proportional to $V^\perp_{\q}$ contribute to $K^{\tn{eff}}_{\tn{xx,exc}}$, such that
\beq
K_{\tn{xx},\tn{exc}}^{\tn{eff}} = - \bigg\langle\sum_q V^\perp_{\q} 2 q_y^2 \left(\overline{\rho}_{\q ,\uparrow}  \overline{\rho}_{-\q ,\downarrow} + \overline{\rho}_{\q ,\downarrow}  \overline{\rho}_{-\q ,\uparrow}\right)\bigg\rangle.
\eeq

As we had noted earlier, evaluating the expectation value requires an actual knowledge of the many-body state of interest. In the present problem, one can evaluate the correlation function with respect to a many-body state, $|\Psi\rangle =  \prod_{\k} \frac{1}{\sqrt{2}}(c_{\k,\uparrow}^\dag + c_{\k,\downarrow}^\dag)|0\rangle $, constructed out of the vacuum, $|0\rangle$. $|\Psi\rangle$ serves as a good trial wavefunction (it is the ground state in the isotropic limit where $V_{\q,\uparrow} = V_{\q, \downarrow} = V_{\q}^\perp$). The above simplifies to,
\beq
\langle \Psi | K_{\tn{xx},\tn{exc}}^{\tn{eff}} |\Psi \rangle &=& \frac{1}{A^2}\sum_{\q} V^\perp_{\q}  q_y^2 \sum_{\k_1, \k_2} \delta_{\k_1,\k_2 + \q} e^{-|q|^2/2 + i \q\wedge \k_1 /2 - i \q\wedge \k_2 /2}\\
&=& \frac{1}{8 \pi^2}\int dq  ~ V^\perp_{q}  q^3 e^{-q^2/2},
\label{eq:D_QH_bilayer}
\eeq
where $N/A = \nu_{\tn{tot}} / 2\pi$. Since $\chi^{\tn{eff}}_{\tn{xx},\tn{exc}}(q_x = 0, q_y \rightarrow 0 ,\omega = 0)$ vanishes, the resulting expression for $D^{\tn{eff}}_{s,\tn{exc}}$ agrees with the well-known result for the exciton phase-stiffness in Ref.\cite{moon95}. Going beyond the above example, we can apply the same formalism to compute (bounds on) the excitonic phase-stiffness for generic interacting flat-band models, using more complicated variational wavefunctions.

\section{Solvable limit for lowest Landau-level-like wave functions}
\label{sec:ll}

The discussion in the previous section, especially in the context of quantum Hall systems, illustrates the utility of nearly solvable many-body wavefunctions associated with lowest Landau level physics. Inspired by this, we return to the problem of superconductivity, but this time around consider the features that arise in the LLL rather than generic flat-bands. Specifically, we will analyze the effect of a uniform Berry curvature vs. fluctuations thereof on the integrated optical spectral weight, and relatedly the SC phase-stiffness.  

\subsection{Uniform Berry curvature}

To begin with, let us consider a completely flat topological band with Chern number, $|C|=1$, whose wave functions are given by those of the LLL. We begin with a single active LLL for spinless electrons. The low energy (projected) Hamiltonian is: $\hi = \sum_\q V_\q \overline{\rho}_{\q} \overline{\rho}_{-\q}$, where $\overline\rho_\q$ is the projected density operator with form factors $\lambda(\k,\q) \equiv \langle u_\k| u_{\k-\q}\rangle$. For a single active band, the covariant derivative defined in Eq.~\ref{eq:cov_D_k} can be simplified to,
\begin{equation}
\mathcal{D}_x\left[\lambda(\k,\q)\right] \equiv \partial_{k_x}\lambda(\k,\q) + \left( i \mathcal{A}_{k,x} -  i \mathcal{A}_{k-q,x}  \right) \lambda(\k,\q),
\end{equation}
where $\mathcal{A}_{k,\mu} = -i \langle u_\k| \partial_{k_\mu} u_\k\rangle$ is the Berry connection as before, and $\mathcal{D}_x$ is related to the projected position operator by
\begin{equation}
    \left[ \mathbb{P}\hat{X}\mathbb{P}, \overline{\rho}_\q\right] = \sum_\k c_\k^\dag c^{\phantom\dagger}_{\k-\q} \mathcal{D}_x\left[\lambda(\k,\q)\right].
\end{equation}
Let $|u_{\k,0}\rangle$ be the magnetic Bloch wave function for the LLL for the usual spatially uniform magnetic field and $\lambda_0(\k,\q)$ be the corresponding form factor. We then have $\mathcal{D}_x\left[\lambda_0(\k,\q)\right] = i q_y \lambda_0(\k,\q)$. This immediately implies that $[\mathbb{P}\hat{X}\mathbb{P}, H_{\tn{int}}] = 0$, such that the corresponding $\kxx = 0$. Note that this is a consequence of perfect translational symmetry and the emergent dipole conservation in the LLL due to the relationship between momentum and dipole moment in LLL. (See \cite{murthy03Hamiltonian,wang16half} and the references therein for a discussion of the dipole picture in the LLL.)

Note that a vanishing $\kxx$ for the single LLL in the clean limit with projected interactions (and the cyclotron-gap set to infinity) immediately implies that $\tn{Re} [\sigma_{\tn{xx}}(\omega)]$ vanishes for all $\omega$; see Eq.~\ref{eq:partial_f_sum}. Moreover, the above statement regarding the longitudinal response applies to any filling of the LLL, independent of the specific low-energy description, including e.g., the composite Fermi liquid at $\nu=1/2$ \cite{hlr,sonPRX,shi2022gifts}. Our observation is in line with Kohn's theorem \cite{kohn1961}; the cyclotron resonance frequency (unaffected by interaction)  saturates the $f-$sum rule, with no sub-cyclotron frequency spectral weight in the $q\rightarrow0$ limit. Relatedly, since the external uniform electric field only couples to the center-of-mass(CoM) motion \cite{kohn1961,yip89} while the LLL dynamics preserves the CoM motion due to the conservation of total momentum and dipole moment, there is no intra-Landau level absorption. Recently, Ref.~\cite{lake2022dipolar} has also demonstrated a vanishing optical conductivity at all frequency for dipole-conserving systems by considering the commutator between the dipole moment operator and the Hamiltonian. However, to reiterate, the discussion does not include the effects of broken continuous translation symmetry by disorder, lattice potential, or sample boundaries. Moreover, as we show next, including additional copies of LLL carrying distinct quantum numbers immediately leads to the generation of a finite $\kxx$.

Let us now enlarge the low-energy Hilbert space and study a model with spinful electrons. Consider a system with overall time-reversal symmetry such that the spin up (down) electrons occupy LLL with Chern numbers $+1$ ($-1$). We are interested in a density-density interaction that couples these degrees of freedom as $H_{\tn{int}} = \sum_\q V_\q \bar{\rho}_{\q,\uparrow} \bar{\rho}_{-\q, \downarrow}$. The action of the covariant derivatives on the form factors for the $\pm$ bands are given by,
\beq
 \mathcal{D}_{k_x} \left[ \langle u_{\k,\uparrow}| u_{\k-\q,\uparrow}\rangle\right]  = -i q_y  e^{-|q|^2/4 + i \q\wedge \k /2}, \mathcal{D}_{k_x} \left[ \langle u_{\k,\downarrow}| u_{\k-\q,\downarrow}\rangle\right]  = i q_y  e^{-|q|^2/4 - i \q\wedge \k /2}.
\eeq
In general, this leads to a non-vanishing contribution to the diamagnetic response, 
\begin{equation}
    \kxx = - \frac{\mathcal{B}^2}{A}\sum_{\q} \langle \rho_{\q,\uparrow} \rho_{-\q,\downarrow}\rangle 4 q_y ^2 V_\q,
    \label{eq:D_pm_chern}
\end{equation}
where $\mathcal{B} = \frac{C a^2}{2\pi}$ is the strength of the uniform Berry curvature (with $C = 1$), and the lattice constant $a$ is set to $\sqrt{2\pi}$. Note the similarity of the above setup with that of the model analyzed in  Sec.~\ref{sec:qmc} based on Ref.~\cite{hofmann2020superconductivity}. The latter is a lattice generalization of this setup with a non-uniform distribution of the Berry curvature; we shall return to this discussion in the next subsection.

We can estimate the expectation value in Eq.~\ref{eq:D_pm_chern} by considering a good variational wavefunction, rather than bounding the many-body expectation value simply based on the filling of the band. Let us consider the system to be at half-filling (i.e. at total filling $\nu = 1$) with a uniform pairing state $|\Psi\rangle =  \prod_{\k} (u_\k + v_\k c_{\k,\uparrow}^\dag c_{-\k,\downarrow}^\dag )|0\rangle$ as a proxy for the ground state, where $|0\rangle$ is the vacuum and $u_\k,~v_\k$ are (complex) variational parameters satisfying $|u_\k|^2 + |v_\k|^2 = 1$. We then have,
\beq
    \langle \Psi |\kxx|\Psi\rangle &=& \frac{-\mathcal{B}^2}{A}\sum_{\q} V_{\q} 4 q_y^2 \sum_{\k_1, \k_2} \delta_{\k_1,-\k_2 } u_{\k_1-\q}^* v_{\k_1-\q}  v_{\k_1}^* u_{\k_1} e^{-|q|^2/2 + i \q\wedge \k_1 /2 + i \q\wedge \k_2 /2}.\nn\\
\eeq
Since the summation over $\k_1$ and $\k_2$ is bounded from above by the case when $|u_\k| = |v_\k| = 1/\sqrt{2}$ with a uniform pairing phase in momentum space,
\beq    
 \langle \Psi |\kxx|\Psi\rangle \leq \frac{\mathcal{B}^2}{8 \pi^2}\int dq ~  |V_{\q}|  q^3 e^{-q^2/2}.
\eeq
The similarity to the result for the exciton stiffness in the bilayer quantum Hall problem in Eq.~\ref{eq:D_QH_bilayer} is not a complete coincidence. These two systems can actually be related by a partial particle-hole transformation associated with one of the spin species \cite{NB93,sodemann2017composite,bi2021excitonic}. 

\subsection{Non-uniform Berry curvature} 

Based on the result obtained for a single LLL, it is natural to analyze next the effect of an inhomogeneous distribution of Berry curvature in momentum space. To study this within the framework of LLL, we shall now include the effect of a spatially periodic magnetic field. The wave functions for such LLL in a periodic magnetic field have been studied in the past \cite{dubrovin1980ground,haldane2018origin}, and have seen a resurgence of interest more recently  \cite{ledwith2020fractional,wang2021exact}. 

The form factor for this particular class of Bloch wave functions can be written as $\lambda(\k,\q) = \lambda_0(\k,\q) g(\k,\q)$. Here $\lambda_0(\k,\q)$ is the form factor for LLL wavefunction in a uniform magnetic field as before, and $g(\k,\q)$ is periodic in $\k$ with $g(\k,\q) = g(\k+\vec{G},\q)$, with $\vec{G}$ a reciprocal lattice vector and $g(\k,0) = 1$. The explicit form of $g(\k,\q)$ will be specified below, and is determined by the spatial fluctuation associated with the magnetic field, or equivalently the fluctuations of the Berry curvature in momentum space. The action of the covariant derivative is then given by,
\begin{equation}
    \mathcal{D}_x \left[\lambda(\k,\q)\right] = i q_y \lambda(\k,\q)  + d_x\left[g(\k,\q)\right] \lambda(\k,\q),
    \label{eq:form_fac_LLL_periodic}
\end{equation}
where the first term comes from $\mathcal{D}_x \left[\lambda_0(\k,\q)\right]$ and we define $d_x\left[g(\k,\q)\right]$ as
\begin{equation}
\begin{split}
 d_x\left[g(\k,\q)\right] \equiv ~\partial_{k_x} \ln\left[g(\k,\q)\right]- \lim_{\alpha\rightarrow 0} \partial_{\alpha_x} g(\k,\vec\alpha) +  \lim_{\alpha\rightarrow 0} \partial_{\alpha_x} g(\k-\q,\vec\alpha), 
\end{split}
\end{equation}
where the first term takes into account the contribution from the momentum derivative in $g(\k,\q)$ and the second and third terms take into account the change in Berry connections from $g(\k,\q)$.
Therefore, the effective diamagnetic response can be expressed as,
\begin{equation}
    \begin{split}
        \kxx = &\sum_{\q,\k_1,\k_2} V_\q \la c_{\k_1}^\dag c_{\k_1-\q}^{\phantom\dag} c_{\k_2}^\dag c_{\k_2+\q}^{\phantom\dag}\ra~ \lambda(\k_1,\q) \lambda(\k_2,-\q)\\
        &\times \Big[\left\{d_x\left[g(\k_1,\q)\right] + d_x\left[g(\k_2,-\q)\right]\right\}^2\\&+ \partial_{k_{1,x}}\left\{d_x\left[g(\k_1,\q)\right]\right\} +\partial_{k_{2,x}}\left\{ d_x\left[g(\k_2,-\q)\right]\right\} \Big].
    \end{split}
    \label{eq:LLL_Kxx}
\end{equation}
Now we study a specific example of a periodic magnetic field; see App.~\ref{app:LLL} and Ref.~\cite{dubrovin1980ground} for details. For a small fluctuation of the magnetic field around the uniform background value, 
\begin{equation}
    \begin{split}
        B(x,y) \approx 1+ \pi \phi_1 \left[\cos(\sqrt{2\pi} x) + \cos(\sqrt{2\pi} y)\right],
    \end{split}
\end{equation}
where the lattice constant is set to $\sqrt{2\pi}$ and $0\leq\phi_1<1$ characterizes the strength of the spatially periodic part of the magnetic field, thereby controlling the fluctuation of the Berry curvature. Given the form factor (App.~\ref{app:LLL}), and in the same small $\phi_1$ limit, the Berry curvature is given by,
\begin{equation}
    \mathcal{B}_{\k} \approx 1-\pi \phi_1 e^{-\frac{\pi}{2}} \left[\cos(\sqrt{2\pi} k_x) + \cos(\sqrt{2\pi} k_y)\right].
\end{equation}
In this limit, since $g(\k,\q) \approx O(\phi_1)$, we only retain the contribution from the last line in Eq.~\ref{eq:LLL_Kxx} and find an upper bound,
\begin{equation}
\begin{split}
  D_s \leq & C(\nu) 2 \sqrt{2} \pi |\phi_1| e^{-\frac{\pi}{2}} \sum_{\q} |V_\q| e^{-\frac{|q|^2}{2}} \cosh^2\bigg(\frac{\sqrt{2\pi} q_y}{4}\bigg).
\end{split}
\end{equation}
Therefore, the above computation illustrates that while the uniform Berry curvature distribution in the LLL limit does not contribute to $\kxx$, the latter is bounded by the fluctuation of the Berry curvature once its distribution becomes non-uniform in momentum-space. Interestingly, in the single LLL problem, while both the Berry curvature and the quantum metric are non-vanishing (and uniformly distributed in momentum-space), $\kxx$ vanishes. 

Let us finally consider the analogous problem with an enlarged low-energy Hilbert space by including spinful electrons. Just as in the problem in the previous section, we assume that the spin up and down electrons occupy LLL with opposite Berry curvature. 

Let $\langle u_{\k,\uparrow}| u_{\k-\q,\uparrow}\rangle = \lambda(\k,\q)$ and $\langle u_{\k,\downarrow}| u_{\k-\q,\downarrow}\rangle = \lambda(\k-\q,-\q)$, where $\lambda(\k,\q)$ is the form factor of LLL under periodic magnetic field in Eq.~\ref{eq:form_fac_LLL_periodic}. The choice of the form factors is to let $B_{k,\downarrow} + B_{k,\uparrow}=0$.
Therefore, the diamagnetic response can be written as,
\begin{equation}
    \begin{split}
         \kxx  = &\sum_{\q,\k_1,\k_2} V_\q \la c_{\k_1,\uparrow}^\dag c_{\k_1-\q,\uparrow}^{\phantom\dag} c_{\k_2,\downarrow}^\dag c_{\k_2+\q,\downarrow}^{\phantom\dag}\ra~ \lambda(\k_1,\q) \lambda(\k_2+\q,\q)\\
        &\times \Big[\left\{2i q_y + d_x\left[g(\k_1,\q)\right] + d_x\left[g(\k_2+\q,\q)\right]\right\}^2\\&+ \partial_{k_{1,x}}\left\{d_x\left[g(\k_1,\q)\right]\right\} +\partial_{k_{2,x}}\left\{ d_x\left[g(\k_2+\q,\q)\right]\right\} \Big].
    \end{split}
    \label{eq:LLL_Kxx_pmChern}
\end{equation}
In order to evaluate the correlators, we make a simplifying assumption and consider a uniform pairing state, $|\Psi\rangle =  \prod_{\k} (u_\k + v_\k c_{\k,\uparrow}^\dag c_{-\k,\downarrow}^\dag )|0\rangle$, where $u_\k = v_\k = \frac{1}{\sqrt{2}}$. Expanding around $\phi_1$, we have,
\begin{equation}
    \begin{split}
        \langle \Psi |\kxx|\Psi\rangle =& \frac{1}{4\pi^2 A}\sum_{\q,\k_1} V_{\q} u_{\k_1-\q}^* v_{\k_1-\q}  v_{\k_1}^* u_{\k_1} e^{-|q|^2/2} \left[ -4 q_y^2 \right.\\
        & - \frac{\phi_1}{2} \sum_j e^{-\frac\pi 2 + i \k_j \wedge \k_1}(4 q_y^2 + k_{j,y}^2) \left(e^{-\frac{q \bar{k}_j}{2}} + e^{\frac{\bar{q} k_j}{2}}-1 - e^{i \q \wedge \k_j}\right)\\
        & \left.-i q_y \phi_1 \sum_j e^{-\frac\pi 2+ i \k_j \wedge \k_1}  \left(i k_{j,y}(e^{-\frac{q \bar{k}_j}{2}}- e^{\frac{\bar{q} k_j}{2}})- k_{j,x} +k_{j,x} e^{i \q \wedge \k_j}\right)\right]  + O(\phi_1^2).
        \label{spinfulChernNU}
    \end{split}
\end{equation}
 It is readily seen that the $O(\phi_1)$ term vanishes due to the integration over $\k_1$ and the leading contribution arises at $O(\phi_1^2)$. Denoting the difference in $\langle \Psi |\kxx|\Psi\rangle$ evaluated for the LLL with a non-uniform vs. uniform Berry curvature distribution as $\Delta \langle \Psi |\kxx|\Psi\rangle$, we obtain $\Delta \langle \Psi |\kxx|\Psi\rangle \approx c_1 v \phi_1^2$ for a potential $V_\q = v/q$ ($c_1\sim O(1)$ number); see App.~\ref{app:LLL}. Thus, the contribution to $\kxx$ increases with the fluctuation of the Berry curvature distribution. However, this conclusion is not meant to suggest that a non-uniform distribution of Berry curvature is essential for supporting superconductivity in flat-bands; explicit computations for trivial flat-bands with an identically vanishing Berry curvature are known to have a finite $T_c$ \cite{hofmann2022superconductivity} and a finite $\kxx$ \cite{DMDC1}.


\section{Discussion}
\label{sec:BCS}

Flat-band superconductivity, in spite of being a seemingly old problem \cite{Shaginyan1990,Kopnin2011,Volovik2013}, continues to reveal a number of new and fascinating puzzles that lie beyond any weak-coupling description. The mechanism leading to a superconducting ground state, if any, for the class of generic isolated flat-bands that we analyze here in the presence of projected density-density interactions remains poorly understood. In the absence of a detailed knowledge of the many-body ground state for the interacting problem, here we have analyzed upper bounds on the superconducting, and a closely related excitonic superfluid, phase-stiffness without invoking any unjustified mean-field approximations.   

The past few years have brought to the forefront the important role played by band-topology, as encoded in the Berry curvature and Chern number, and band-geometry, as determined by the Fubini-Study metric, on the superfluid stiffness in flat-band models. In particular, the analyses based on applications of BCS mean-field theory highlight the important role played by the geometry associated with the flat-band Bloch wavefunctions in delocalizing the Cooper pairs \cite{Torma15,Bernevig19,Torma19,Rossi19,Bernevig_review}; the latter ingredient has been shown to be at least one of the relevant factors based on more sophisticated numerically exact treatments of the interacting flat-band problem \cite{hofmann2020superconductivity,peri2021fragile,hofmann2022superconductivity,Bernevig21}. However, interactions in the same flat-band limit can drive a number of competing instabilities \cite{hofmann2020superconductivity,hofmann2022superconductivity}, which makes it difficult to make generic statements about the nature of the ground state in this non-perturbative limit.

In this work, we have taken a complementary approach, where we place reasonably tight upper-bounds on the superfluid phase-stiffness for density-density interactions projected to generic flat-bands, and simultaneously highlighting a number of subtleties associated with the electromagnetic response of the resulting low-energy theory. These include a special attention to the non-commuting limits of $\omega \rightarrow 0$ and the transverse $\q \rightarrow 0$, and constructing the correct ``projected'' gauge transformation asosociated with only the low-energy degrees of freedom.

We can use our framework to compute the effective low-energy diamagnetic response (and relatedly phase-stiffness) if the ground-state many-body wavefunction is known exactly, or if a good variational wavefunction can be constructed. We have illustrated this for both the excitonic and the superconducting phase-stiffness for the lowest Landau level using various generalizations of the bilayer quantum Hall problem. When the ground state wavefunctions are unknown, we can still use our framework to put conservative upper bounds on the largest possible magnitude of the same quantity. We have illustrated this for two different sign-problem-free models of interacting flat-bands, where the value for the phase-stiffness is known from independent quantum Monte-Carlo computations. Depending on the microscopic details associated with the form of the Bloch wavefunctions, our estimated upper bounds usually turn out to be approximately a factor of $O(10)$ of the known value from QMC. 

Moving forward, it is natural to apply our framework to realistic narrow-band moir\'e materials exhibiting superconductivity (and excitonic attraction), and compare to the wealth of experimental data. It would be interesting to analyze the implications of the difference between the actual $T_c$ and the approximate upper bounds on the underlying phenomenology. Moreover, when the gap to the remote bands ($\Delta$) is not infinity, as is the case for realistic materials, it would be interesting to incorporate their $O(1/\Delta)$ contribution to the electromagnetic response in a systematic fashion. Finally, developing the analogous theory for upper bounds on the optical spectral weight starting with projected electron-phonon interactions remains an important open problem.

\begin{center} {\bf Acknowledgements} \end{center}
We thank E. Berg, B.A. Bernevig, J. Hofmann, S. Kivelson, J.F. Mendez-Valderrama, G. Murthy, M. Randeria, C. Repellin, E. Tulipman and J. Wang for insightful discussions. We thank E. Tulipman for feedback on an earlier version of this manuscript. D.M. is supported by a Bethe/KIC postdoctoral fellowship at Cornell University. D.C. is supported by a faculty startup grant at Cornell University. D.C. dedicates this paper to the memory of Prof. Amit Dutta. 

\begin{appendix}
\section{Lowest Landau level in a periodic magnetic field}\label{app:LLL}

In this appendix we provide a brief overview of the LLL wave functions under a periodic magnetic field following closely Refs.~\cite{dubrovin1980ground, mao2021strongly}. Let us consider a wave function $\psi(\vec{x})$ satisfying the following equation,
\begin{equation}
    (\Pi_x+i\Pi_y)\psi(\vec{x}) = 0,
    \label{eq:psi}
\end{equation}
where $\Vec{\Pi} = \Vec{P}-\Vec{A}$ is the kinetic momentum (where we set all fundamental constants to 1). We are interested in the form of $\psi(\vec{x})$ for the LLL under a non-uniform magnetic field with the Hamiltonian $H = (\Pi_x-i\Pi_y)(\Pi_x+i\Pi_y)$; we express $\psi(\vec{x}) = e^{-\varphi(\vec{x})}f(z)$, where $f(z)$ is a holomorphic function of $z=x+iy$ and $(\partial_x^2 + \partial_y^2)\varphi(\vec{x}) = B(\vec{x})$. Note that for a uniform magnetic field $B$, $\psi(\vec{x}) = e^{-|z|^2/4}f(z)$. In the following discussion, we consider a periodic magnetic field with the symmetry of a square lattice enclosing $2\pi$ magnetic flux per unit cell. 

Note that the Hamiltonian commutes with the discrete magnetic translations $\hat{T}_1 : \psi(x,y) \rightarrow \psi(x+\sqrt{2\pi},y) e^{-i\sqrt{\frac{\pi}{2}}y}$ and $\hat{T}_2 : \psi(x,y) \rightarrow \psi(x,y+\sqrt{2\pi}) e^{i\sqrt{\frac{\pi}{2}}x}$, where $\sqrt{2\pi}$ is the lattice constant.  We can construct the magnetic Bloch wave function $\psi_{\vec{k}}(\vec{x})$ as an eigenstate of the magnetic translations $\hat{T}_1$ and $\hat{T}_2$ with eigenvalues $e^{i k_x \sqrt{2\pi}}$ and $e^{i k_y \sqrt{2\pi}}$, respectively. The analytical form of $\psi_{\vec{k}}(\vec{x})$ is given by,
\begin{equation}
    \psi_{\vec{k}}(\vec{x}) = \sigma(z-a_1) e^{a z-\varphi(\vec{x})},
    \label{eq:Bloch}
\end{equation}
where $a = \frac12 \left(k_y - \sqrt{\pi/2} + i k_x - i \sqrt{\pi/2}\right)$, $a_1 = k_y - \sqrt{\pi/2} - i k_x + i \sqrt{\pi/2}$, and $\sigma(z)$ is the Weierstrass sigma function with the same periodicity as the periodic magnetic field; see Ref.~\cite{dubrovin1980ground} for details. To be concrete, let us consider a specific $\varphi(\vec{x})$,
\begin{equation}
    \varphi(\vec{x}) =  \frac{|z|^2}{4} - \frac{1}{2}\log\left[ 2+ 2 \phi_1 \left(\cos(\sqrt{2\pi} x)+\cos(\sqrt{2\pi} y\right)\right],
\end{equation}
where $0\leq \phi_1< 1$ characterizes the strength of the periodic part of the magnetic field. The associated magnetic field is then given by,
\begin{equation}
    B(\vec{x}) = 1 + \pi \phi_1 \frac{2\phi_1 + 2\phi_1 \cos(\sqrt{2\pi} x) \cos(\sqrt{2\pi} y) +  \cos(\sqrt{2\pi} x)+\cos(\sqrt{2\pi} y)}{\left\{1+  \phi_1 \left[\cos(\sqrt{2\pi} x)+\cos(\sqrt{2\pi} y)\right]\right\}^2}.
    \label{eq:mag_field}
\end{equation}

The Bloch function is defined as $u_{\vec{k}}(\vec{x}) = \frac{\psi_{\vec{k}}(\vec{x})}{\sqrt{\langle \psi_{\vec{k}}|\psi_{\vec{k}}\rangle}} e^{-i \vec{k} \cdot \vec{x}}$ and the form factor $\lambda(\vec{k},\vec{q}) \equiv \langle u_{\vec{k}}|u_{\vec{k}-\vec{q}}\rangle $ can therefore be written as,
\begin{equation}
    \begin{split}
         \lambda(\vec{k},\vec{q}) &= e^{-\frac{|q|^2}{4}+ i \frac{q \bar{a}_1+ \bar{q} a_1}{4}} \frac{1- \frac{\phi_1}{2} \sum_j e^{-\frac{\pi}{2}+ i \frac{k_j \bar{a}_1+ \bar{k}_j a_1}{2}-\frac{q \bar{k}_j}{2}}}{\sqrt{\bigg(1-\frac{\phi_1}{2} \sum_j e^{-\frac{\pi}{2}+ i \frac{k_j \bar{a}_1+ \bar{k}_j a_1}{2}}\bigg)\bigg(1-\frac{\phi_1}{2}\sum_j e^{-\frac{\pi}{2}+ i \frac{k_j(\bar{a}_1-i \bar{q})+ \bar{k}_j (a_1+i q)}{2}}\bigg)}}\\
        &\equiv \lambda_0 (\vec{k},\vec{q}) g(\vec{k},\vec{q}).
    \end{split}
    \label{eq:form_factor_LLL}
\end{equation} 
The derivatives relevant for Eq.~\ref{eq:LLL_Kxx} are then given by,
 \begin{equation}
     \begin{split}
         d_x\left[g(\vec{k},\vec{q}) \right] = &\frac{\frac{\phi_1}{2}\sum_j i k_{j,y}e^{-\frac{\pi}{2}+ i \frac{k_j \bar{a}_1+ \bar{k}_j a_1}{2}-\frac{q \bar{k}_j}{2}}}{1- \frac{\phi_1}{2} \sum_j e^{-\frac{\pi}{2}+ i \frac{k_j \bar{a}_1+ \bar{k}_j a_1}{2}-\frac{q \bar{k}_j}{2}}} \\
         &-\frac{1}{2}\frac{\frac{\phi_1}{2} \sum_j k_j  e^{-\frac{\pi}{2}+ i \frac{k_j \bar{a}_1+ \bar{k}_j a_1}{2}}}{1-\frac{\phi_1}{2} \sum_j e^{-\frac{\pi}{2}+ i \frac{k_j \bar{a}_1+ \bar{k}_j a_1}{2}}}+\frac{1}{2}\frac{\frac{\phi_1}{2}\sum_j \bar{k}_j e^{-\frac{\pi}{2}+ i \frac{k_j (\bar{a}_1-i \bar{q})+ \bar{k}_j (a_1+i q)}{2}}}{1-\frac{\phi_1}{2}\sum_j e^{-\frac{\pi}{2}+ i \frac{k_j (\bar{a}_1-i \bar{q})+ \bar{k}_j (a_1+i q)}{2}}}\\
         \approx &\frac{\phi_1}{2}\sum_j e^{-\frac{\pi}{2}+ i \frac{k_j \bar{a}_1+ \bar{k}_j a_1}{2}} \left[i k_{j,y} e^{-\frac{q \bar{k}_j}{2}} - \frac{k_j}{2} + \frac{\bar{k}_j}{2} e^{-\frac{q \bar{k}_j}{2} + \frac{\bar{q}k_j}{2}} \right],
     \end{split}
 \end{equation}
and
\begin{equation}
    \begin{split}
        &\partial_{k_x} d_x\left[g(\vec{k},\vec{q}) \right]  \approx \frac{\phi_1}{2}\sum_j e^{-\frac{\pi}{2}+ i \frac{k_j \bar{a}_1+ \bar{k}_j a_1}{2}} k_{j,y}^2 \left[ e^{-\frac{q \bar{k}_j}{2}} - \frac{1}{2} -\frac{1}{2} e^{-\frac{q \bar{k}_j}{2} + \frac{\bar{q}k_j}{2}} \right]\\
        &=\pi \phi_1 e^{-\frac{\pi}{2}} \left[ \cos(\sqrt{2\pi} k_x) +  \cos\left(\sqrt{2\pi} (k_x-q_x)\right)\right.\\&\left. - 2\cosh(\frac{\sqrt{2\pi} q_y}{2}) \cos\left(\sqrt{2\pi} (k_x-\frac{q_x}{2})\right) - 2i \sinh(\frac{\sqrt{2\pi} q_y}{2}) \sin\left(\sqrt{2\pi} (k_x-\frac{q_x}{2})\right) \right].
    \end{split}
\end{equation}
Since $|\cos|$ and $|\sin|$ are bounded by one,
\begin{equation}
    \begin{split}
        \left|\partial_{k_x} d_x\left[g(\vec{k},\vec{q}) \right]\right| \leq 4 \sqrt{2} \pi |\phi_1| e^{-\frac{\pi}{2}} \cosh^2\bigg(\frac{\sqrt{2\pi} q_y}{4}\bigg)
    \end{split}
\end{equation}

For the analogous spinful case with $\pm$ Chern bands having opposite Berry curvature, we noted in Eq.~\ref{spinfulChernNU} that $ \langle \Psi |\kxx|\Psi\rangle$ vanishes at $O(\phi_1)$. At $O(\phi_1^2)$ we obtain, 
\begin{equation}
    \begin{split}
        &\Delta \langle \Psi |\kxx|\Psi\rangle\\
        = &\frac{1}{8\pi^3}\int d^2 q V_\q \left\{- 2 q_y^2 \phi_1^2 e^{-\pi-|q|^2/2} \left[4- 4 \cosh{\frac{\sqrt{2\pi} q_x}{2}} \cos{\frac{\sqrt{2\pi} q_y}{2}} - 4 \cosh{\frac{\sqrt{2\pi} q_y}{2}} \cos{\frac{\sqrt{2\pi} q_x}{2}}\right. \right.\\
        &+\left. \cos{\sqrt{2\pi} q_x} + \cos{\sqrt{2\pi} q_y} + \cosh{\sqrt{2\pi} q_x} + \cosh{\sqrt{2\pi} q_y}\right]\\
        &+ \left. q_y \phi_1^2 e^{-\pi-|q|^2/2} \sum_j \left[( k_{j,y} e^{-\frac{\bar{q} k_j}{2}} - i \frac{\bar{k}_j}{2} + i \frac{k_j}{2} e^{i \k_j\wedge \q})(-1 + e^{-i \k_j \wedge \q} + 2e^{- \frac{\q\cdot \k_j}{2}}\cos{\frac{\q\wedge\k_j}{2}})+c.c\right]\right\},
    \end{split}
\end{equation}
which can be simplified with $V_\q=v/q$ to yield the result quoted in the main text.
\end{appendix}

\bibliography{references.bib}

\end{document}